\documentclass[aps,twocolumn,nofootinbib,preprintnumbers,superscriptaddress]{revtex4-2}

\usepackage{color}
\usepackage{bbm}
\usepackage{tikz}

\newcommand{\nb}[1]{\color{blue}}

\newcommand{\hl}[1]{\color{magenta}}

\usepackage[
pagebackref=false,
colorlinks=true,
linkcolor=blue,
urlcolor=blue,
filecolor=black,
citecolor=red,
pdfstartview=FitV,
pdftitle={},
pdfauthor={},
pdfsubject={},
pdfkeywords={},
pdfpagemode=None,
bookmarksopen=true
]{hyperref}

\usepackage[normalem]{ulem}
\usepackage{amsmath, amssymb, bm, amsfonts, mathtools, braket}
\usepackage{enumerate}
\usepackage{amsfonts}
\usepackage{epsfig}
\usepackage{mathbbol}
\usepackage{mathrsfs}
\usepackage{braket}
\usepackage{ulem}

\usepackage[utf8]{inputenc}
\DeclareUnicodeCharacter{0304}{\ensuremath{\bar{}}}

\begin{document}

\title{Non-Hermitian boost deformation}

\date{\today}

\author{Taozhi Guo}
\thanks{The authors are listed in alphabetical order.}
\affiliation{Department of Physics, Princeton University, Princeton, New Jersey 08544, USA}

\author{Kohei Kawabata}
\thanks{The authors are listed in alphabetical order.}
\affiliation{Department of Physics, Princeton University, Princeton, New Jersey 08544, USA}
\affiliation{Institute for Solid State Physics, University of Tokyo, Kashiwa, Chiba 277-8581, Japan}

\author{Ryota Nakai}
\thanks{The authors are listed in alphabetical order.}
\affiliation{Department of Physics, Kyushu University, Fukuoka, 819-0395, Japan}

\author{Shinsei Ryu}
\thanks{The authors are listed in alphabetical order.}
\affiliation{Department of Physics, Princeton University, Princeton, New Jersey 08544, USA}

\begin{abstract}
The Hatano-Nelson model is one of the most prototypical non-Hermitian models that exhibit the intrinsic non-Hermitian topological phases and the concomitant skin effect.
These 
phenomena 
unique to non-Hermitian topological systems
originate from the Galilean transformation.
Here, we extend such an idea to a broader range of systems based on an imaginary boost deformation and identify the corresponding energy-twisted boundary conditions. 
This imaginary boost deformation complexifies spectral parameters of integrable models and can be implemented by the coordinate Bethe ansatz. 
We apply the imaginary boost deformation to several typical integrable models, including free fermions, the Calogero-Sutherland model, and the XXZ model. 
We find the complex-spectral winding in free fermion models under the periodic boundary conditions and the non-Hermitian skin effect under the open boundary conditions. 
The interaction effect is also shown in the two-particle spectrum of the XXZ model. 
\end{abstract}

\maketitle
\tableofcontents

\section{Introduction}

%
%
%
%
%

Recently, non-Hermitian physics has attracted extensive interest as it describes open systems interacting with the environment~\cite{Konotop-review, Christodoulides-review, Bergholtz-review}. 
As the energy spectrum becomes complex in non-Hermitian systems, interesting physics that cannot be seen in traditional Hermitian systems emerges. 
As a prime example, non-Hermitian systems can support two types of complex-energy gaps, line gap and point gap~\cite{Shen_Zhen_Fu_2018, Gong_2018, Kawabata_2019}.
Specifically, point-gapped non-Hermitian systems can host topological phases that have no analogs in Hermitian systems.
Such intrinsic non-Hermitian topological phases also lead to a new type of bulk-boundary correspondence called non-Hermitian skin effect~\cite{Lee_2016, MartinezAlvarez_2018, Yao_Wang_2018, Kunst_2018, Lee_Thomale_2019, Yokomizo_Murakami_2019, Zhang_2020, Okuma_2020}, in which an extensive number of eigenstates are localized at the boundaries.

The Hatano-Nelson model~\cite{Hatano_Nelson_1996,Hatano_Nelson_1997} is one of the most 
fundamental 
non-Hermitian models as it provides interesting physics in a fairly simple fashion. 
Originally, it was introduced to describe the movements of the vortex lines in superconductors, while similar non-Hermitian models were recently realized in single photons~\cite{Xiao-19-skin-exp} and ultracold atoms~\cite{liang-22}, as well as a variety of open classical systems~\cite{Brandenbourger-19-skin-exp, Helbig-19-skin-exp, Weidemann-20-skin-exp, Palacios-21, Zhang-21}.
In the Hatano-Nelson model, non-Hermiticity is implemented as an imaginary-valued gauge field that makes the hopping amplitudes nonreciprocal.
This model is solvable using a similarity transformation that reduces the wave functions to the Hermitian counterparts, which captures the skin effect and 
Anderson transition
induced by non-Hermiticity.
Even in the presence of many-body interactions, the skin effect still survives~\cite{Fukui-98Nucl, Fukui-98PRB, Lee_Lee_Yang_2020, Zhang_Chen_2020, Liu_2020, Xu_2020, Shen_Lee_2022, Alsallom_2022, Zhang_Neupert_2022, Kawabata_Shiozaki_Ryu_2022}.  
Interestingly, Hatano and Nelson also argue that the imaginary gauge field can be understood as 
a boost of Galilean transformation.
However, such a relationship between non-Hermiticity and the Galilean 
boost
has been discussed only for 
simple free fermionic models,
and the general implications of this approach have been unclear.

For an integrable system, there is a set of integrable deformations that 
keeps the infinitely many commuting conserved charges (quantities), such as boost deformation~\cite{Bargheer_2009,10.21468/SciPostPhys.8.2.016}. 
With boost deformation, the boost parameter associated with conserved charges is generated by the boost operator. Interestingly, the boost deformation modifies the spectrum but keeps the energy spectrum real. 
Previously, such a boost transformation, as well as the concomitant energy-twisted boundary conditions, was applied to thermal quantum transport~\cite{Nakai_Guo_Ryu_2022}.

In this work, 
by generalizing the boost parameter to an imaginary parameter, 
we obtain a complex energy spectrum with such a boost deformation and study non-Hermitian physics. 
Put differently, we seek an analog to the Hatano-Nelson approach and extend the idea of Galilean transformation to a broader context 
that is applicable to more generic models.
While the original discussion by Hatano and Nelson focused on noninteracting models, our approach can be applied to many-body interacting models.
We consider 
transformation of a Hermitian Hamiltonian that 
rescales coordinates and
leads to a non-Hermitian Hamiltonian. 
Such a rescale transformation can also be understood as putting the system in a boosted reference frame in spacetime coordinates. 
For lattice systems, especially integrable systems, the boost transformation can be implemented as an imaginary boost deformation of conserved charges. 
In the language of the algebraic Bethe ansatz (i.e., algebraic formulation of the Bethe ansatz based on transfer matrices), such a boost deformation complexifies spectral parameters as it generates an imaginary component for each conserved charge. 
To seek a boundary condition description of such a bulk deformation, we consider 
the U(1) gauge field as it can also be formulated as a phase twisted boundary condition.
Similarly, for the bulk imaginary boost deformation, the boundary transformation can be formulated as what we call the energy-twisted boundary condition~\cite{Nakai_Guo_Ryu_2022}.

With such a transformation, an originally Hermitian Hamiltonian becomes a non-Hermitian Hamiltonian with a complex spectrum. 
We show that the imaginary boost deformation induces the 
topological properties 
unique to non-Hermitian systems
and the concomitant skin effect.
With the tools of the Bethe ansatz for integrable models, we 
study not only free fermions but also 
interacting models where the interplay between many-body interactions and the boost deformation is worth investigating. 

The rest of this work is formulated as follows. 
In Sec.~\ref{sec:dirac_fermion}, we formulate the energy-twisted boundary conditions of a chiral Dirac fermion and calculate its deformed spectrum. 
We also introduce the rescale transformation that leads to a non-Hermitian Hamiltonian. 
In Sec.~\ref{sec:ABA}, we discuss the algebraic Bethe ansatz formalism and implement the imaginary boost deformation. 
In Sec.~\ref{sec:free_Fermion}, we study the boost deformation of free fermions and obtain their complex spectra.
We confirm the complex-spectral winding under the periodic boundary conditions and the consequent skin effect under the open boundary conditions.
We also present the results of two-particle energy spectra and their winding numbers. 
In Sec.~\ref{sec:CS}, we study the imaginary boost deformation of the Calogero-Sutherland model as a prototypical example of interacting integrable models. 
In Sec.~\ref{sec:XXZ}, we investigate the imaginary boost deformed XXZ model. 
Finally, we conclude in Sec.~\ref{sec:conclusion} and discuss further directions.

\section{
Dirac fermion} \label{sec:dirac_fermion}


The boost deformation for integrable lattice systems is defined in terms of the flow equation~\cite{10.21468/SciPostPhys.8.2.016,Bargheer_2009}, 
\begin{align}
&\frac{dH(\kappa)}{d\kappa}=i[B[H(\kappa)],H(\kappa)],
    \label{eq:realboostdeformation}
\end{align}
where $\kappa \in \mathbb{R}$ is the boost parameter, and $B[H(\kappa)] = \sum_x xh_x,$ is the boost operator with $H = \sum_x h_x$. 
Such a deformation keeps the Hamiltonian Hermitian and integrable. 
In this work, we generalize the boost deformation 
from real to imaginary
as 
\begin{align}
&\frac{dH(\kappa)}{d\kappa}=[B[H(\kappa)],H(\kappa)].
    \label{eq:boostdeformation}
\end{align}
The boost operator is well-defined for an infinitely long system.
We will later comment on the boost deformation for finite 
systems. 

As a warm-up, we consider
a (1+1)D chiral Dirac fermion 
on an infinitely long system,
\begin{align}
  {H} =
  \int dx\,
  {\psi}^{\dag}
  \mathcal{H}
  {\psi},
  \quad
  \mathcal{H} = - i v \partial_x,
  \label{eq:chiraldiracfermion}
\end{align}
where $\psi(x)$ is a fermion field operator and $v$ is the Fermi velocity. 
In this system, 
\begin{align}
{H}(\kappa) &=
  \int dx\,
  {\psi}^{\dag}
  \mathcal{H}(\kappa)
  {\psi}, \\
B[H(\kappa)]
&=
\int dx\,
  {\psi}^{\dag}
  x\mathcal{H}(\kappa)
  {\psi},
\end{align}
are the deformed Hamiltonian and the boost operator, respectively.
The solution under the initial condition in Eq.~(\ref{eq:chiraldiracfermion}) is
\begin{align}
H(\kappa)
=
\int dx\,
  {\psi}^{\dag}
  \frac{-iv\partial_x}{1-iv\kappa} 
  {\psi}.
\end{align}
For a finite system of length $L$, the system is considered 
to be subject to
 the periodic boundary conditions
 $\psi \left( x+L \right) = \psi \left( x \right)$, and the single-particle wave functions are given as $f_k \left( x \right) = e^{ikx}/\sqrt{L}$ with $k = 2\pi n/L$ ($n\in\mathbb{Z}$).
The corresponding single-particle eigenenergies are 
$\varepsilon_k = vk/(1-iv\kappa)$.

The same spectral deformation can be obtained by imposing a boundary condition twisted by the energy~\cite{Nakai_Guo_Ryu_2022},
\begin{equation}
    \psi(x+L) = e^{\kappa L H} \psi(x) e^{-\kappa L H},
\end{equation} 
which is referred to as the energy-twisted boundary condition, while the Hamiltonian remains unchanged.
For the imaginary boost deformation, the energy-twisted boundary conditions read,
\begin{equation}
    f_{k}(x+L) = e^{ikL}f_{k}(x) = e^{-\kappa\varepsilon_k L}f_{k}(x).
\end{equation}
Then, the momentum is solved by
\begin{equation}
   ( k - i\kappa\varepsilon_k ) L = 2\pi n,\quad\text{i.e.,}\quad k  = \frac{2\pi n}{L} \frac{1}{1-i v\kappa}
        \label{eq: chiral-eTBC-momentum}
\end{equation}
with $n\in \mathbb{Z}$. \
The energy twisted boundary condition is equivalent to the imaginary boost deformation as the momentum quantization is the same. 
The bulk deformation gives non-Hermitian Hamiltonians, which correspond to the nonunitary boundary twist operator $e^{\kappa LH}$. 

The thermal (imaginary-time) partition function of the (1+1)D chiral Dirac fermion on a torus reads,
\begin{equation}
    Z = \mathrm{Tr}\,e^{-\beta H} = \mathrm{Tr}\,q^{L_0 - c/
    24}
\end{equation}
with $q \coloneqq e^{2\pi i\tau}$ and modular parameter $\tau \coloneqq i\beta v/L$.
Under the energy-twisted boundary conditions, the modular parameter is modified to
\begin{equation}
    \tau \rightarrow \frac{i\beta v}{L} \frac{1}{1-iv\kappa}.
\end{equation}

The mode expansion of the fermion operator is given by $\psi(x) = \sum_k b_k e^{ikx}$. To make a connection to the energy-twisted boundary conditions, we consider a new reference frame that rescales the original frame such that 
\begin{equation}
x' = \frac{x}{1-iv\kappa}.
\end{equation}
Then, in the new $x'$ reference frame, the mode expansion becomes
\begin{equation}
    \psi(x') = \sum_{k} b_k e^{ikx'(1-iv\kappa)}.
\end{equation}
If we require the periodic boundary conditions in the $x'$ reference frame, i.e., $\psi(x') = \psi(x'+L)$, we get 
Eq.~(\ref{eq: chiral-eTBC-momentum}).
Therefore, in the new reference frame, the Hamiltonian is non-Hermitian 
because of the complex momentum.
In this sense, the energy-twisted boundary conditions can be understood as a generalization of the imaginary gauge transformation introduced by Hatano and Nelson~\cite{Hatano_Nelson_1996, Hatano_Nelson_1997}.
There, the Galilean boost $x' = x - vt = x +igt/m$ for a free particle $\varepsilon = p^2/2m$ leads to the non-Hermitian Hamiltonian $\varepsilon = (p+ig)^2/2m$.
Notably, the Hatano-Nelson model is a prototypical model that exhibits the complex-spectral winding~\cite{Shen_Zhen_Fu_2018, Gong_2018, Kawabata_2019} and the concomitant skin effect~\cite{Lee_2016, MartinezAlvarez_2018, Yao_Wang_2018, Kunst_2018, Lee_Thomale_2019, Yokomizo_Murakami_2019, Zhang_2020, Okuma_2020}, both of which are topological phenomena inherent in non-Hermitian systems.
Similarly, we find that the imaginary boost deformation leads to a wide variety of complex-spectral winding and skin effect, as shown in the following for several models.
In contrast to the original Hatano-Nelson model, the boost deformation gives rise to energy-dependent gauge fields and longer-range hoppings, which make properties of integrable models richer.

\section{Boost deformation of integrable models}
    \label{sec:ABA}

Before investigating several models, we here 
review the boost deformation of integrable models and
demonstrate that our imaginary boost deformation is generally equivalent to complexifying spectral parameters of integrable models.
For a one-dimensional integrable model on a lattice, there exist Lax operators $L_{an}(\lambda)$ where $a$ labels the auxiliary vector space, $n$ labels the local Hilbert space, and $\lambda$ is the spectral parameter associated with the auxiliary vector space. 
The monodromy matrix is defined as 
 \begin{equation}
     T_a(\lambda) = L_{aN}(\lambda) L_{aN-1}(\lambda) \cdots L_{a1}(\lambda)
 \end{equation}
with the RTT relation
\begin{equation}
    R_{12}(\lambda_1 - \lambda_2) T_1(\lambda_1) T_2(\lambda_2) =  T_2(\lambda_2)T_1(\lambda_1)R_{12}(\lambda_1 - \lambda_2),
\end{equation}
where $R_{12}(\lambda_1 - \lambda_2)$ is the $R$ matrix of the integrable model. 
The transfer matrix is defined as 
\begin{equation}
    T(\lambda) = \text{Tr}_a{T_a(\lambda)} .
\end{equation}
With the transfer matrix, the conserved operators 
are generated as
\begin{equation}
   I_n (\lambda) = (-i)^{n+1}\frac{d^n}{d\lambda^n} \log{T(\lambda) } . 
\end{equation}
Here, $I_1$ is the Hamiltonian $H =I_1(\lambda = 0)$,
and eigenvalues of $I_0$ are momenta.
Following Refs.~\cite{THACKER1986348,1982JETP...55..306T,10.1143/PTP.69.431}, we define the boost operator
\begin{equation}
    B = \sum_x xh_x,
\end{equation}
where $h_x$ is the local Hamiltonian satisfying $H = \sum_x h_x$. Such a boost operator generates the Lorentz boost in the sense that it boosts the rapidity of the transfer matrix by
\begin{equation}
    \frac{d T(\lambda)}{d\lambda} = [B,T(\lambda)].
\end{equation}
Then, with this boost operator, we generate higher-order conserved charges by
\begin{equation}
    [B,I_n] = iI_{n+1}.
        \label{eq:charge boost}
\end{equation}

Generalizing the Hermitian deformation, we consider the non-Hermitian boost deformation 
as 
\begin{equation}
    \frac{d I_n^{\kappa}(\lambda)}{d\kappa} = [B, I_n^{\kappa}(\lambda)] = iI_{n+1}^{\kappa}(\lambda), 
    \label{eq:imaginary deformation}
\end{equation}
with $n>0$ and the boost parameter $\kappa$.
We notice here that the boost operators generate non-Hermitian higher-order charges such as the Hamiltonian. 
In this sense, the spectral parameter of the transfer matrix is complexified by the deformation $\lambda \rightarrow \lambda(\kappa)$. 
As such, we call such a deformation generated by Eq.~\eqref{eq:imaginary deformation} the imaginary boost deformation, and the Hamiltonian becomes non-Hermitian after this deformation. 
In comparison with Eq.~(\ref{eq: chiral-eTBC-momentum}),
this can be identified as a bulk description of a   boundary twist similar to the case of the phase twisted boundary condition that represents the bulk U(1) gauge field. 

As shown in Ref.~\cite{10.21468/SciPostPhys.8.2.016}, if the state is deformed as 
\begin{equation}
    \frac{d|\psi_a^{\kappa}\rangle}{d\kappa} = B|\psi_a^{\kappa}\rangle,
    \label{eq:similaritytransformation_integrable}
\end{equation}
the eigenvalues of the conserved operators $ I_n^{\kappa }|\psi^{\kappa} \rangle = q_{n}^{\kappa }  |\psi^{\kappa} \rangle$ are unchanged, i.e., 
\begin{equation}
    \frac{d q_n^{\kappa}}{d\kappa} = 0 \quad \left( n>0 \right).
\end{equation}
In this case, the momentum $P^{\kappa}$ and energy $E^{\kappa}$,
 which are respectively the eigenvalues of $I_0^{\kappa}$ and $I_1^{\kappa}$, are related by
\begin{equation}
P^{\kappa} = P^{\kappa = 0} +i\kappa E^{\kappa = 0}.
    \label{eq:momentum boost}
\end{equation}
For each quasiparticle, we have $P^{\kappa} = \sum_{j}p^{\kappa}_j$ and $p^{\kappa}_j = p^{\kappa = 0 } + i\kappa \varepsilon_j^{\kappa = 0}$. 
With this formalism, the momentum quantization of quasiparticles is solved using the coordinate Bethe ansatz as 
\begin{equation}
  \label{twisted BAE}
  e^{ i (p^{\kappa}(v_j) - i\kappa \varepsilon_j) L}
  = \prod_{k(\neq j)} S(v_j-v_k),
\end{equation}
where $v_j$ is the rapidity of the quasiparticle $j$ and $S(v_j - v_k)$ is the scattering phase between quasiparticles. 

As an example, we consider the XXX Heisenberg model~\cite{Katsura_2010, Murg_2012}
\begin{equation}
    H = \sum_{i = 1}^L \left[ \sigma_i^x \sigma_{i+1}^x + \sigma_i^y \sigma_{i+1}^y + (\sigma_i^z \sigma_{i+1}^z -1) \right].
\end{equation}
The Lax operator of this model is 
\begin{equation}
    L(\lambda) = \begin{pmatrix}
1 & 0 & 0 & 0 \\
0 & b(\lambda) & c(\lambda) & 0 \\
0 & c(\lambda) & b(\lambda) & 0 \\
0 & 0 & 0 & 1
\end{pmatrix}
\end{equation}
with $b(\lambda) = \frac{\lambda - \frac{i}{2}}{\lambda + \frac{i}{2}}$ and $c(\lambda) = \frac{i
}{\lambda + \frac{i}{2}}$. 
The momentum of each particle is given by $ e^{-ip} = b(\lambda)$. 
For the imaginary boost deformation, the spectral parameter is changed to $\lambda^0 \rightarrow \lambda(\kappa)$, and the momentum is changed to $p^{\kappa} = p^{\kappa = 0} + i\kappa \varepsilon_p^{\kappa = 0}$. 
Therefore, we write the boosted momentum as 
\begin{equation}
    e^{-ip^{\kappa }} = \frac{\lambda(\kappa) - \frac{i}{2}}{\lambda(\kappa)+  \frac{i}{2}} = e^{-i(p^0+i\kappa \varepsilon_p^0)}.
\end{equation}
With the identification $e^{-ip^{\kappa = 0}} = b(\lambda^0) $ and $\varepsilon_p^{\kappa = 0} = b(\lambda^0) + 1/b(\lambda^0) - 2$,
we get
\begin{align}
\lambda(\kappa) =  &\frac{i}{2}\frac{(2 i \lambda^0 -1) e^{-4\kappa \varepsilon_p^0 }+(2 i \lambda^0
   +1)}{(2 i \lambda^0 -1)  e^{-4\kappa \varepsilon_p^0}-(2 i \lambda^0
   +1)} \nonumber \\
   =& -\frac{i}{2} \frac{2 i \lambda^0 \cosh{(2\kappa \varepsilon_p^0)} +  \sinh{(2\kappa \varepsilon_p^0)}}{2 i \lambda^0 \sinh{(2\kappa \varepsilon_p^0)} +  \cosh{(2\kappa \varepsilon_p^0)}},
\end{align}
which indicates that the spectral parameter is complexified under the imaginary boost deformation.



\section{
Free fermion}
    \label{sec:free_Fermion}

\subsection{Free fermion in free space}
    \label{subsec: free fermion in free space}

\begin{figure}[t]
 \centering
 \includegraphics[width=0.49\linewidth]{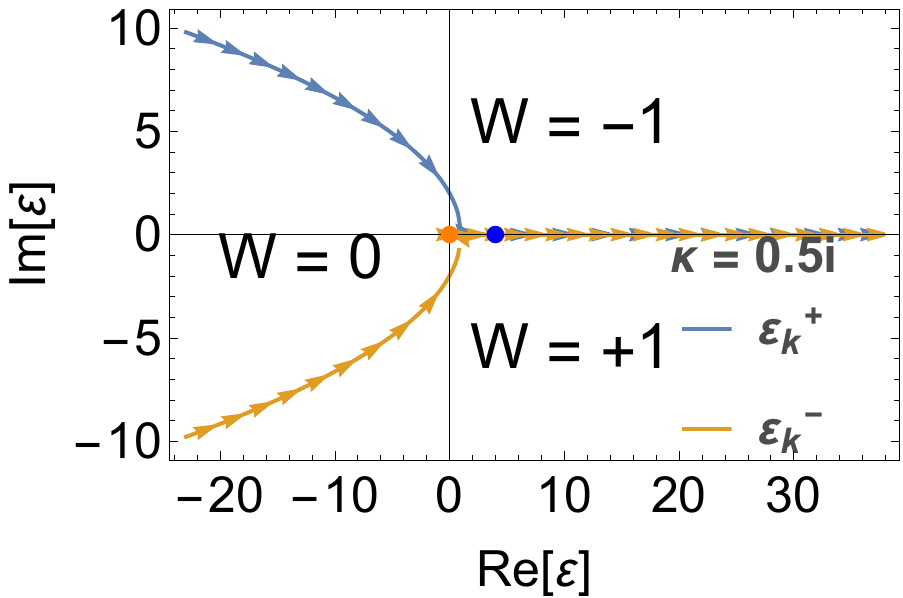}
 \includegraphics[width=0.49\linewidth]{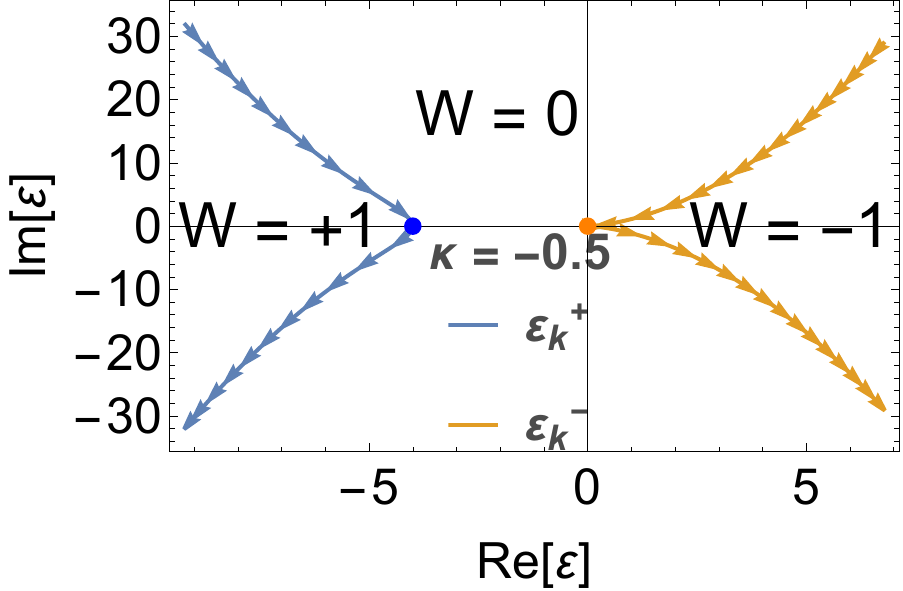}
 \caption{Complex energy spectrum of free fermion $\varepsilon_k = k^2$ for (left)~real boost $\kappa = 0.5i$ and (right)~imaginary boost $\kappa = -0.5 $. 
 The orange and blue dots are for $k_0 = 0$. 
 The arrows on the curves indicate the flow directions of momentum from $k_0 = -\infty$ to $k_0 = \infty$.
 The complex-spectral winding number is denoted by $W$.
 }
    \label{fig:CS single particle}
\end{figure}

\subsubsection{Periodic boundary conditions}

We consider the free fermion in free space with the quadratic dispersion 
\begin{equation}
    \varepsilon_k = k^2.
        \label{eq: k2-dispersion}
\end{equation}
Under the periodic boundary conditions $f \left( 0 \right) = f \left( L \right)$, the momentum satisfies $k_0 = 2\pi n/L$ with $n \in \mathbb{Z}$. 
Under the 
energy-twisted 
boundary conditions, 
momentum $k$ and energy $\varepsilon_k$ satisfy
\begin{equation}
    k - i\kappa \varepsilon_k = k_0
        \label{eq: eTBC - k2 dispersion}
\end{equation}
with the boost parameter $\kappa$.
When the non-Hermitian term is a constant, the model reduces to the Hatano-Nelson model~\cite{Hatano_Nelson_1996, Hatano_Nelson_1997}.
In our boosted model, however, the non-Hermitian term $i\kappa \varepsilon_k$ is no longer a constant and depends on energy.
Combining Eqs.~(\ref{eq: k2-dispersion}) and (\ref{eq: eTBC - k2 dispersion}), we have
\begin{equation}
    \varepsilon_k = k^2  = (k_0 + i\kappa \varepsilon_k)^2,
        \label{eq: k2 - boost eq}
\end{equation}
leading to
\begin{equation}
    \varepsilon_k^{\pm} = -\frac{1}{2\kappa^2}(1 - 2i k_0\kappa \pm \sqrt{1 -  4ik_0\kappa}).
    \label{eq:deformedspectrum_continuousfreefermion}
\end{equation}
Notably, the two branches of energy appear in the presence of the boost deformation, leading to the rich behavior of complex spectra even for free fermion.
In the zero-boost limit $\kappa \to 0$, we have
\begin{align}
    \varepsilon_k^{+} \simeq -\frac{1}{\kappa^2} \to \pm\infty,\quad
    \varepsilon_k^{-} \to k_0^2.
\end{align}
For $k_0 = 0$, we have
\begin{align}
     \varepsilon_k^{+} = - \frac{1}{\kappa^2},\quad
    \varepsilon_k^{-} = 0.
\end{align}

We consider both real and imaginary boost parameters, as shown in Fig.~\ref{fig:CS single particle}.
The energy spectrum is complex even for a real boost $i\kappa\in \mathbb{R}$.
The real spectrum changes into the complex spectrum at $i k_0 \kappa = 1/4$, around which the complex spectrum exhibits the square-root singularity similarly to exceptional points~\cite{Heiss_EP}.
We find that the complex spectrum under the real boost winds in the complex-energy plane. 
Then, we calculate the complex-spectral winding number $W$ defined as
\begin{align}
    W \coloneqq - \int_{k_0 = -\infty}^{k_0 = +\infty} \frac{dk_0}{2\pi i} \frac{d}{dk} \log \det \left[ H_k - \varepsilon \right]
        \label{eq: 1D winding number}
\end{align}
with a reference energy $\varepsilon \in \mathbb{C}$.
This complex-spectral winding number $W$ gives a topological invariant unique to non-Hermitian systems~\cite{Shen_Zhen_Fu_2018, Gong_2018, Kawabata_2019}.
Note that this topological invariant is always trivial for the real spectrum and can be nontrivial only for the complex spectrum.
Our complex spectrum is divided into three regions, two of which we find are characterized by the nontrivial winding number $W = \pm 1$ (see Fig.~\ref{fig:CS single particle}).
The nontrivial winding number implies a current~\cite{Hatano_Nelson_1996, Hatano_Nelson_1997, Kawabata_2021_TFT} and also the skin effect under the open boundary conditions~\cite{Lee_2016, MartinezAlvarez_2018, Yao_Wang_2018, Kunst_2018, Lee_Thomale_2019, Yokomizo_Murakami_2019, Zhang_2020, Okuma_2020}.
It should be noted that even the real boost makes the spectrum complex although the boosted Hamiltonian always preserves Hermiticity.
This originates from the nonlinear nature of the eigenvalue equation~(\ref{eq: k2 - boost eq}) and may be considered as a manifestation of spontaneous breaking of Hermiticity.

For an imaginary boost, the complex spectrum exhibits different behavior (Fig.~\ref{fig:CS single particle}).
As the imaginary boost is turned on, the original spectrum $\varepsilon_k^{-}$ becomes complex, and a point gap is open [see the orange curve in Fig.~\ref{fig:CS single particle}\,(right)].
On the other hand, the other branch $\varepsilon_k^{+}$ of energy emerges from negative infinity and gets closer to the original branch $\varepsilon_k^{-}$ as the boost parameter increases. 
In contrast to the real boost, the two branches never touch each other.
We also calculate the complex-spectral winding number $W$ in Eq.~(\ref{eq: 1D winding number}), as shown in Fig.~\ref{fig:CS single particle}.
The imaginary boost also leads to the nontrivial winding number, which implies the skin effect under the open boundary conditions;
we indeed find the corresponding skin effect under the open boundary conditions, as discussed below.

\subsubsection{Open boundary conditions}

Next, we consider the free fermion with the imaginary boost under the open boundary conditions. 
Similarly to the previous case, the single-particle eigenequation reads
\begin{equation}
    \left( -i \partial_x -i \kappa \varepsilon \right)^2 f \left( x \right) = \varepsilon f \left( x \right)
        \label{eq: free fermion OBC}
\end{equation}
with the boost parameter $\kappa \in \mathbb{R}$.
Instead of the periodic boundary conditions $f \left( L \right) = f \left( 0 \right)$, we impose the open boundary conditions
\begin{align}
    f \left( 0 \right) = f \left( L \right) = 0.
        \label{eq: free fermion free space OBC}
\end{align}
These boundary conditions indeed correspond to the open boundary conditions for the corresponding lattice model.

The Hatano-Nelson model with open boundaries is solvable via a similarity transformation~\cite{Hatano_Nelson_1996, Hatano_Nelson_1997}.
Even though the imaginary gauge field effectively depends on energy $\varepsilon$ in our boosted system, 
we show that we can still introduce the imaginary gauge transformation and then solve our boosted model.
In fact, let us introduce
\begin{align}
    \tilde{f}_{\varepsilon} \left( x \right) \coloneqq e^{ \kappa \varepsilon x} f \left( x \right),
    \label{eq:similaritytransformation_continuousfreefermion}
\end{align}
which now depends on energy $\varepsilon$
[see Appendix~\ref{appendix: boost} for a derivation from Eq.~(\ref{eq:similaritytransformation_integrable})].
Then, Eq.~(\ref{eq: free fermion OBC}) reduces to
\begin{align}
    \left( -i \partial_x \right)^2 \tilde{f}_{\varepsilon} \left( x \right) = \varepsilon \tilde{f}_\varepsilon \left( x \right)
\end{align}
with the open boundary conditions
\begin{align}
    \tilde{f}_{\varepsilon} \left( 0 \right) = \tilde{f}_{\varepsilon} \left( L \right) = 0.
\end{align}
For arbitrary $\varepsilon$, this eigenvalue problem is readily solved as 
\begin{equation}
    \varepsilon = k^2,\quad
    \tilde{f}_{\varepsilon} \left( x \right) \propto \sin \left( kx \right)
\end{equation}
with momenta $k= n\pi/L$ ($n \in \mathbb{N}$).
Hence, the original eigenvalue problem in Eq.~(\ref{eq: free fermion OBC}) is solved as
\begin{equation}
    \varepsilon = k^2,\quad
    f \left( x \right) \propto e^{-\kappa k^2 x} \sin \left( kx \right).
        \label{eq: free fermion OBC solution}
\end{equation}

Thus, the spectrum is entirely real, and no point gap is open, which contrasts with the complex spectrum under the periodic boundary conditions.
All the eigenstates except for the zero modes with $\varepsilon = 0$ are localized at the left (right) edge in the presence of the imaginary boost $\kappa > 0$ ($\kappa < 0$)---non-Hermitian skin effect.
This is compatible with the complex-spectral winding number under the periodic boundary conditions (see Fig.~\ref{fig:CS single particle}).
The localization length of the skin modes is 
\begin{align}
    \xi = \frac{1}{\left| \kappa \varepsilon \right|} = \frac{1}{\left| \kappa \right| k^2}.
\end{align}
Notably, the complex spectrum under the periodic boundary conditions includes an additional contribution from infinity [see the blue curve in Fig.~\ref{fig:CS single particle}\,(right)], which also exhibits the complex-spectral winding. 
The above skin modes do not correspond to this additional complex-spectral winding number but that arising from the original spectrum [see the orange curve in Fig.~\ref{fig:CS single particle}\,(right)].

\subsection{Free fermion on lattice}
    \label{sec: free fermion on lattice}

\subsubsection{Periodic boundary conditions}

Next, we consider a 
translation-invariant free fermion
on a lattice
and the corresponding boost operator,
\begin{align}
 &H(\kappa)=\sum_{xz}t_{z}(\kappa)c_x^{\dagger}c_{x+z}^{\ },\\
 &B[H(\kappa)]=
 \sum_{xz}\left(x+\frac{z}{2}\right)t_{z}(\kappa)c_x^{\dagger}c_{x+z}^{\ }.
\end{align}
The boost deformation generally keeps quadratic fermionic Hamiltonians quadratic.
The imaginary boost deformation, i.e.,
\begin{align}
 \frac{dH(\kappa)}{d\kappa}=[B[H(\kappa)],H(\kappa)],
\end{align}
reduces to
\begin{align}
 \frac{dt_{z}(\kappa)}{d\kappa}=
 -\frac{z}{2}\sum_w t_{w}(\kappa)t_{z-w}(\kappa).
\end{align}
This equation is solved with an initial condition $t_{z}(\kappa=0)=-e^{i\alpha}\delta_{z,1}-e^{-i\alpha}\delta_{z,-1}$.
We introduce the generating function
\begin{align}
\varepsilon(\kappa, k)= \sum_{z} e^{i zk} t_z(\kappa),
\end{align}
where the energy dispersion of the original lattice fermion model is $\varepsilon(\kappa=0,k)=-2\cos(k+\alpha)$.
Then, the equation reduces to the inviscid Burgers equation
\begin{align}
  i\frac{\partial \varepsilon}{\partial \kappa}
  +
  f \frac{\partial \varepsilon}{\partial k}
  =0,
\end{align}
which has a formal solution
\begin{align}
 \varepsilon \left( \kappa,k \right)
 &= \varepsilon \left( \kappa=0, k+i \kappa \varepsilon \left( \kappa,k \right) 
 \right) \nonumber \\
 &= -2 \cos \left( k +
 \alpha + i\kappa \varepsilon \left( \kappa,k \right) \right).
    \label{eq:imaginaryboost}
\end{align}
Strictly speaking, the boost deformation is applicable only to infinite systems, where the boost operator is defined unambiguously.
Here, we try to relax this condition and consider a lattice fermion model under the periodic boundary conditions by assigning $k= 2\pi n/L$ 
($n = 0, 1, 2, \cdots, L-1$).
We assume that the boost deformation is still described by Eq.~(\ref{eq:imaginaryboost}).
This energy dispersion can be effectively considered as that of the Hatano-Nelson model with the energy-dependent imaginary gauge field $\kappa \varepsilon$.

\begin{figure}[t]
 \centering
 \includegraphics[width=0.23\textwidth]{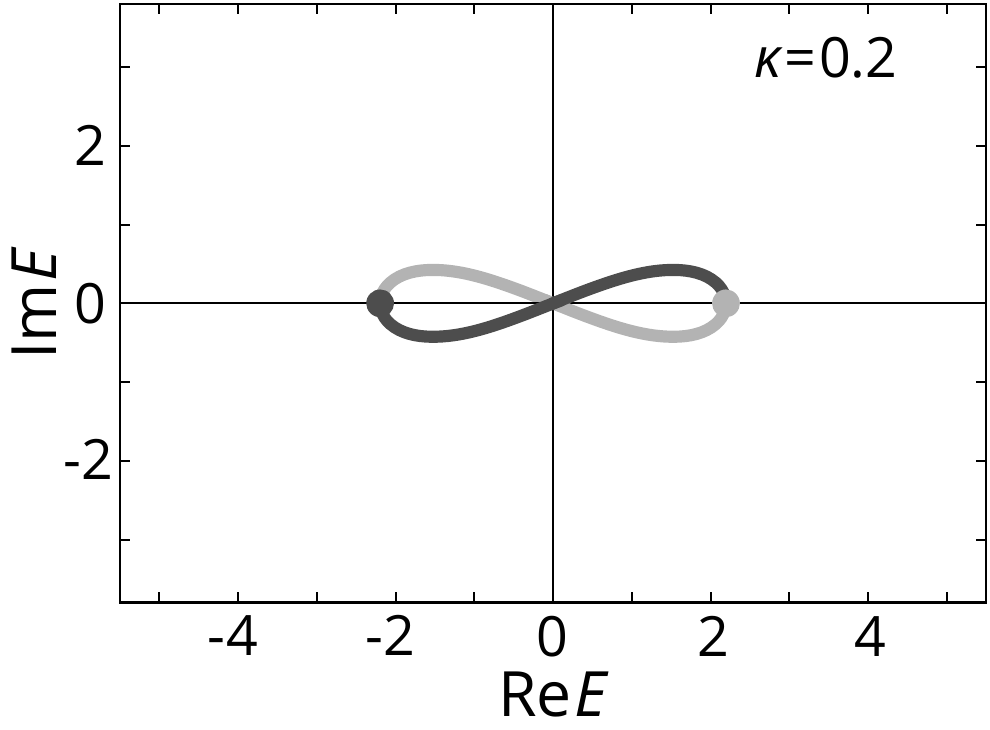}
 \includegraphics[width=0.23\textwidth]{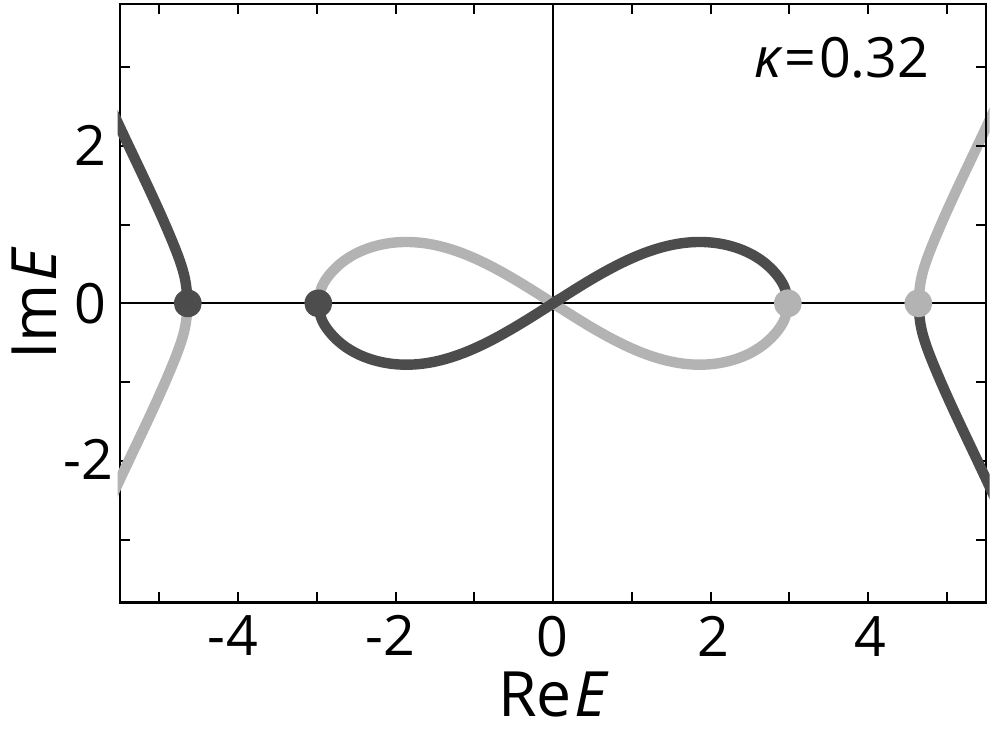}
 \includegraphics[width=0.23\textwidth]{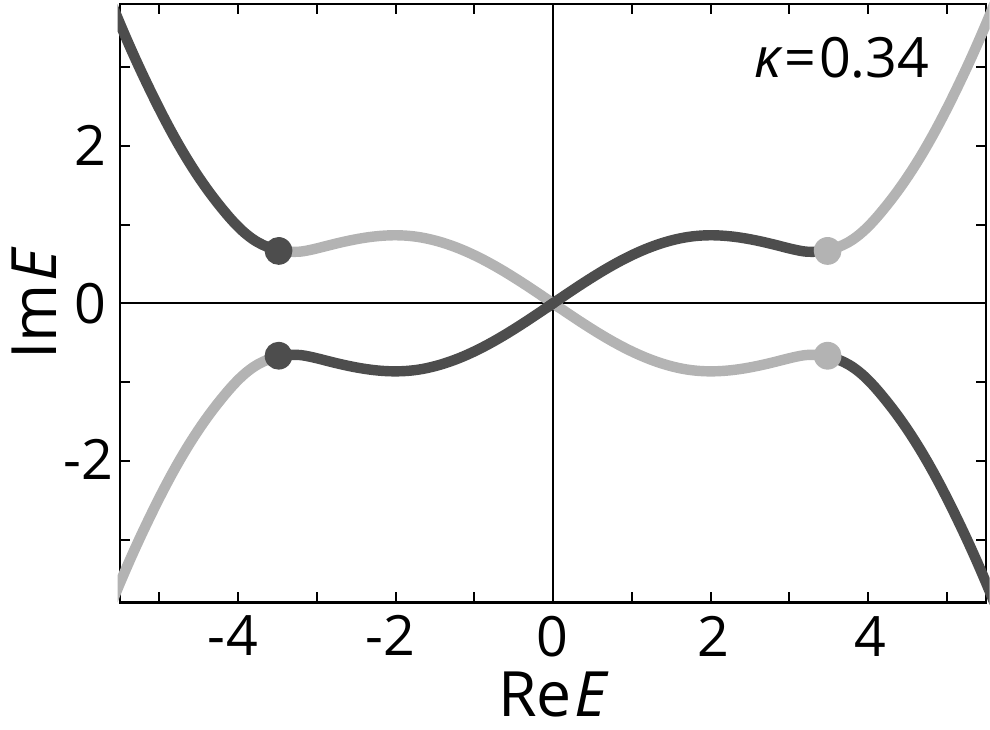}
 \includegraphics[width=0.23\textwidth]{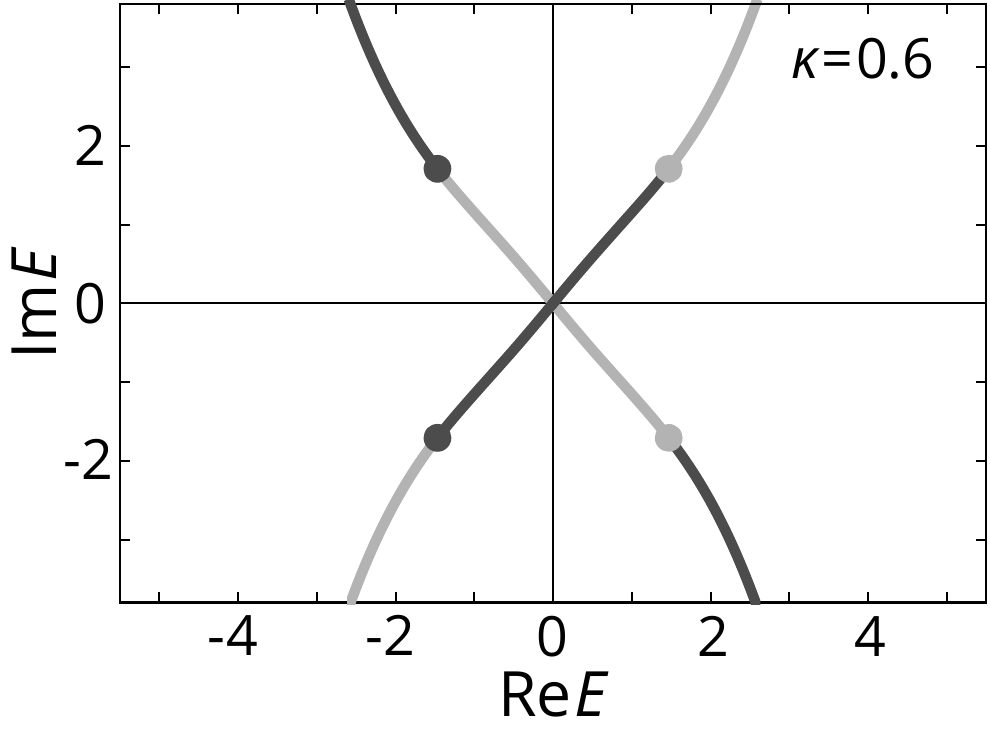}
 \caption{Complex energy spectrum of a free fermion on a lattice under the imaginary boost deformation for (top left)~$\kappa=0.2$, (top right)~$\kappa=0.32$, (bottom left)~$\kappa=0.34$, and (bottom right)~$\kappa=0.6$, respectively.
 The dark-grey points are energies of $k=0$ and the connected lines with the same color are those of $k\in[0,\pi)$, while the light-grey points are of $k=\pi$ and connected lines are of $k\in[\pi,2\pi)$.}
    \label{fig:freefermiondeformation}
\end{figure}

We numerically obtain the complex energy spectrum of a lattice fermion with the imaginary boost, as shown in Fig.~\ref{fig:freefermiondeformation}.
For a small boost, the complex spectrum forms an eight shape.
This eight-shaped complex spectrum can be obtained perturbatively for small energy $\varepsilon$ and boost $\kappa$.
In fact, expanding the boost equation~(\ref{eq:imaginaryboost}) 
for $\left| \varepsilon \right| \ll \left| \kappa \right|^{-1}$ and $\alpha = 0$, 
we have
\begin{align}
    \varepsilon = - 2\cos k - 2i \kappa \sin 2k + O \left( \kappa^2 \right),
\end{align}
which reproduces the eight-shaped complex spectrum numerically obtained in Fig.~\ref{fig:freefermiondeformation}. 
The complex-spectral winding number $W = W \left( \varepsilon \right)$ is also obtained as
\begin{align}
    W \left( \varepsilon \right) = \begin{cases}
    \mathrm{sgn} \left( \kappa \right) & \left( 
    \varepsilon~\text{is inside the left loop} \right); \\
    - \mathrm{sgn} \left( \kappa \right) & \left( 
    \varepsilon~\text{is inside the right loop} \right); \\
    0 & \left( \text{otherwise}\right).
    \end{cases}
        \label{eq: winding number - lattice fermion}
\end{align}
Notably, this result is valid even for the arbitrary boost $\kappa$ as long as the energy $\varepsilon$ is small (i.e., $\left| \varepsilon \right| \ll \left| \kappa \right|^{-1}$).

As we increase the boost parameter $\kappa$, the other pieces of the complex spectra approach the eight-shaped spectrum from infinity.
This situation is similar to the complex spectrum for free fermions in free space, as discussed in Sec.~\ref{subsec: free fermion in free space}.
For $\kappa \gtrsim 0.33$,
we see that the eight-shaped spectrum touches the spectrum coming from infinity and forms a cross-shaped spectrum [see Fig.~\ref{fig:freefermiondeformation} (bottom)].
This spectral phase transition is unlikely to occur in usual non-Hermitian lattice models and originates from the nonlinear nature of the boost equation.
According to our numerical calculations, this spectral phase transition occurs on the real axis in the complex-energy plane. 
Thus, the spectral transition points for $\mathrm{Re}\,\varepsilon > 0$ ($\mathrm{Re}\,\varepsilon < 0$) corresponds to $k=\pi$ ($k=0$).
For $k=\pi$ and $\varepsilon \in \mathbb{R}$, the boost equation~(\ref{eq:imaginaryboost}) 
reduces to
\begin{align}
    \varepsilon = 2 \cosh \kappa \varepsilon.
\end{align}
For the existence of a solution to this equation, $\varepsilon$ is required to satisfy $\varepsilon > 2$.
In this case, we have
\begin{equation}
    \kappa = \frac{\mathrm{arccosh} \left( \varepsilon/2 \right)}{\varepsilon},
\end{equation}
which has two (no) solutions if $\kappa$ is less (larger) than its maximum value.
The spectral transition point corresponds to the maximum value of $\kappa = \kappa \left( \varepsilon \right) = \mathrm{arccosh} \left( \varepsilon/2 \right)/\varepsilon$, i.e.,
\begin{equation}
    \kappa_{\rm c} = 0.331372 \cdots,
        \label{eq: kappa critical}
\end{equation}
which is compatible with the numerical results in Fig.~\ref{fig:freefermiondeformation}.


\subsubsection{Open boundary conditions}
\label{sec: free fermion on lattice OBC}

We also apply the 
imaginary boost deformation to a lattice fermion under the open boundary conditions.
In this case, lattice translation invariance no longer exists and we need to solve the evolution of the hopping amplitude in the matrix form as
\begin{align}
 \frac{dt_{xy}(\kappa)}{d\kappa}=
 \frac{x-y}{2}\sum_{a=1}^L t_{xa}(\kappa)t_{ay}(\kappa).
 \label{eq:boostdeformation_latticefermion}
\end{align}
Assuming the analyticity of the evolution, we expand the hopping amplitude by $\kappa$ as
\begin{align}
 t_{xy}(\kappa)=\sum_{n=0}^{\infty}\frac{\kappa^n}{n!}\frac{d^n t_{xy}(\kappa)}{d\kappa^n}\bigg|_{\kappa=0}.
 \label{eq:deformedhamiltonian_latticefreefermion}
\end{align}
Here, the higher derivatives of the hopping amplitude are evaluated successively via
\begin{align}
 \frac{d^{n+1}t_{xy}(\kappa)}{d\kappa^{n+1}}=
 \frac{x-y}{2}\sum_{a=1}^L \sum_{j=0}^{n}
 \begin{pmatrix}
  n \\ j
 \end{pmatrix}
 \frac{d^{n-j}t_{xa}(\kappa)}{d\kappa^{n-j}}
 \frac{d^{j}t_{ay}(\kappa)}{d\kappa^{j}}.
\end{align}
We calculate the hopping amplitude up to the order of 300 for the system size $L=40$.

\begin{figure}[t]
 \centering
 \includegraphics[width=0.42\textwidth]{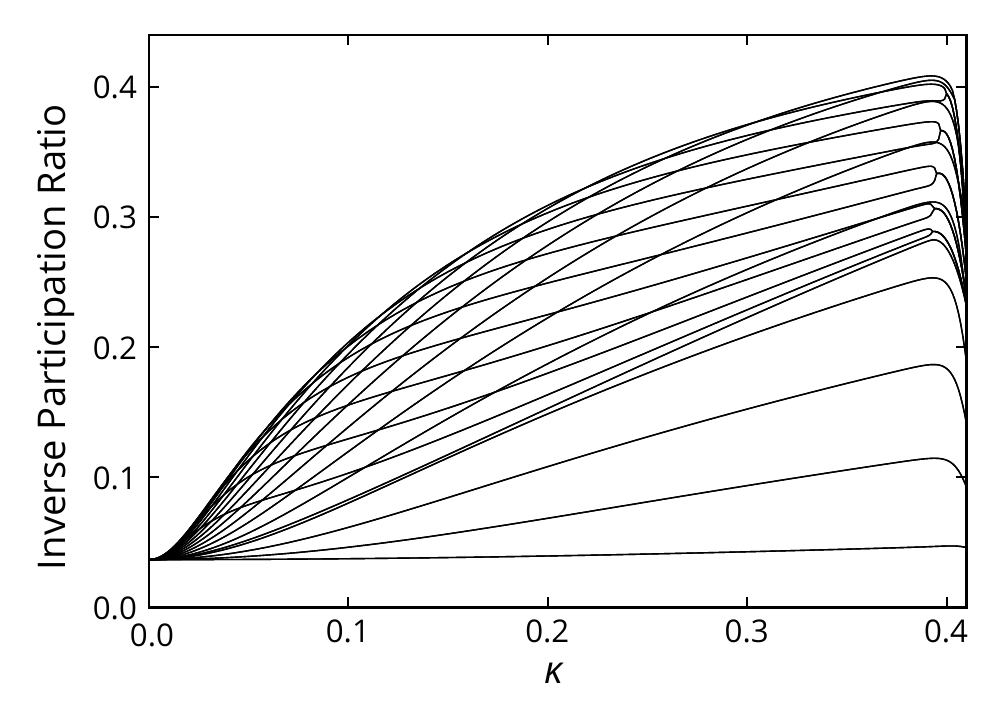}
  \includegraphics[width=0.42\textwidth]{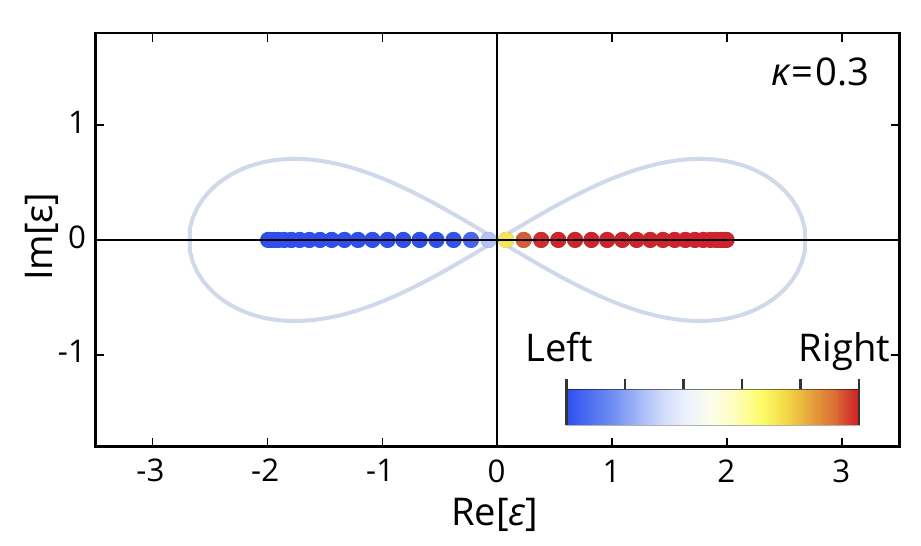}
 \caption{(Top)~Inverse participation ratio (IPR) of all eigenstates as a function of the imaginary boost deformation parameter $\kappa$ for $L=40$. 
 (Bottom)~Complex energy spectrum of a lattice fermion under the open boundary conditions for $\kappa=0.3$.
 The grey solid line is the spectrum under the periodic boundary conditions for reference.
 The color of the open-boundary eigenenergies indicates a weighted probability density in Eq.~(\ref{eq:weightedprobabilitydensity}) quantifying at which side the eigenstates are localized.
 }
    \label{fig:freefermiondeformation_obc}
\end{figure}

We obtain the inverse participation ratio (IPR) $\sum_x|f_x|^4$ of all the eigenstates [Fig.~\ref{fig:freefermiondeformation_obc} (top)], which scales as $O \left( 1/L \right)$ and $O \left( 1 \right)$ for the extended and localized states, respectively.
While the energy spectrum is almost unchanged during the boost deformation, the IPR shows the localization of all the states on the edges except for zero-energy states.
The edges at which localized states are bound are consistent with the winding number $W$ in Eq.~(\ref{eq: winding number - lattice fermion}).
We also show the energy spectrum under the open boundary conditions in Fig.~\ref{fig:freefermiondeformation_obc} (bottom).
The color of the spectrum indicates a weighted probability density 
\begin{align}
 -\sum_{x<(L+1)/2}|f_x|^2+\sum_{x>(L+1)/2}|f_x|^2,
 \label{eq:weightedprobabilitydensity}
\end{align}
which is close to $-1$ ($+1$) when an eigenstate is mostly on the left (right) half of the system.
Clearly, eigenstates with $\varepsilon > 0$ ($\varepsilon < 0$) are localized at the right (left) edge, which is consistent with Eq.~(\ref{eq: winding number - lattice fermion}).
Eigenstates under the open boundary conditions having the winding number $W = \pm 1$ 
typically behave as $f_x \propto e^{\kappa \varepsilon x}$, similarly to Eq.~(\ref{eq: free fermion OBC solution}).

The similarity transformation of the lattice fermion is not as simple as the continuum case (see Appendix~\ref{sec:similaritytransformation_lattice}).
In the continuum case, a local Hamiltonian remains local, at least for the linear and quadratic fermionic models studied in Secs.~\ref{sec:dirac_fermion} and \ref{subsec: free fermion in free space}.
In the lattice case, the boost deformation generates longer-range hoppings, which lead to the highly nonlocal Hamiltonian.

\subsection{Two particles}
    \label{subsec: free fermion - two particle}

\begin{figure}[t]
 \centering
 \includegraphics[width=0.48\linewidth]{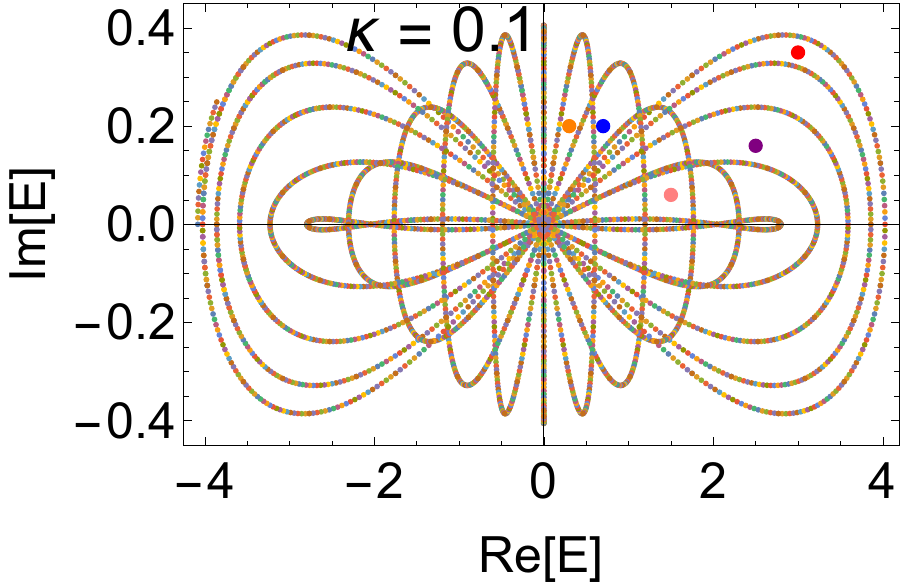}
 \includegraphics[width=0.48\linewidth]{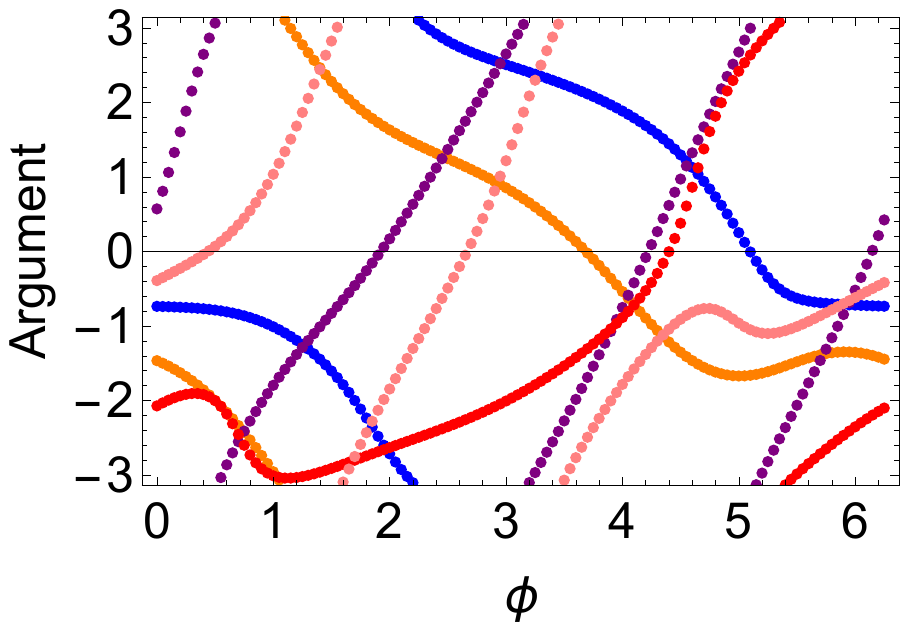}
  \includegraphics[width=0.48\linewidth]{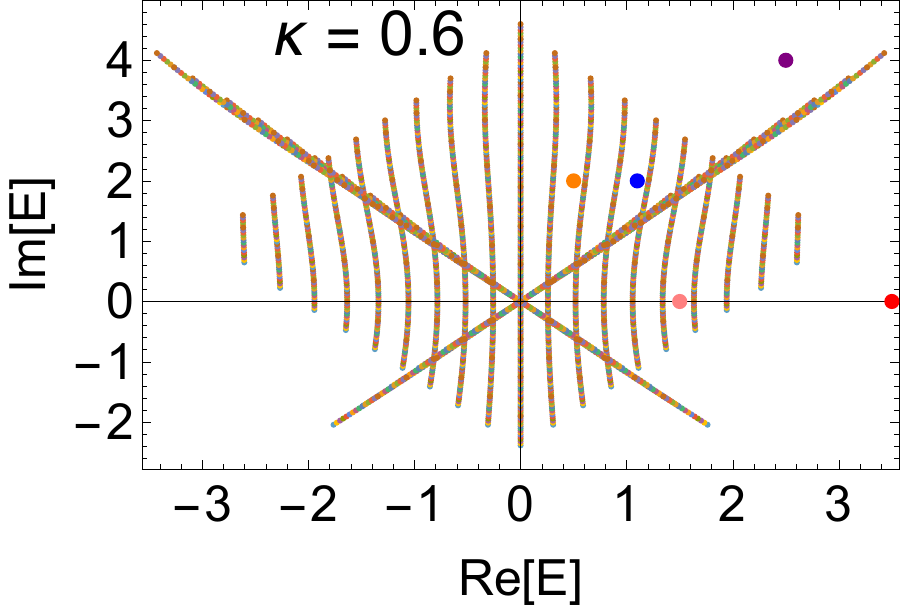}
 \includegraphics[width=0.48\linewidth]{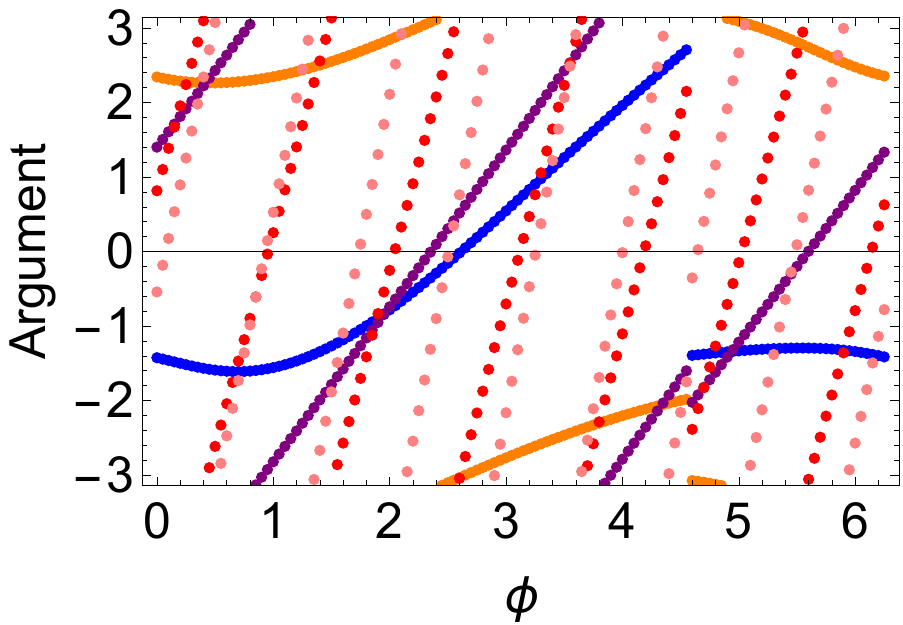}
 \caption{Two-particle complex spectrum and winding number of 
 the free fermion model on a lattice of $L = 20$ with the imaginary boost deformation. 
 The boost parameters are $\kappa = 0.1$ (top) and $\kappa = 0.6$ (bottom). 
 The colored dots on the left plots label the reference energy chosen for the calculations of the winding numbers on the right plot with the same color.}
    \label{fig:twop_freefermion}
\end{figure}

We also study the spectrum in the subspace of two particles for the lattice free fermion under the imaginary boost deformation.
The single-particle boosted spectrum is given by 
\begin{equation}
      \varepsilon(k) = -2\cos\Big(\frac{2\pi}{L}n + \frac{\phi}{
      L} + i\kappa\varepsilon(k) \Big),
\end{equation}
where we introduce the magnetic flux $\phi$ to compute the complex-spectral winding number~\cite{Kawabata_Shiozaki_Ryu_2022}.
Taking two different single-particle energies $\varepsilon_1$ and $\varepsilon_2$ from this single-particle boosted dispersion, we calculate the two-particle spectrum $E = \varepsilon_1 + \varepsilon_2$,
as shown in Fig.~\ref{fig:twop_freefermion}. 
In numerical calculations, it is difficult to capture the complex spectrum coming from infinity.
Here, we only pick the energy branch around the origin and focus on the complex spectrum around there.
For $\kappa < 0.33$, the two-particle energy spectrum forms multiple loops. For $\kappa > 0.33$, on the other hand, the loops expand and open up to lines and do not close. 
This spectral transition coincides with the transition point $\kappa_{\rm c} \simeq 0.33$ in Eq.~(\ref{eq: kappa critical}) for the single-particle energy spectrum. 
For $\kappa = 0.1$ [see Fig.~\ref{fig:twop_freefermion}\,(top)], most loops are smooth but there is one loop $C^*$ close to the 
real axis
that is narrow and crosses itself. 
The loops enclosing $C^*$  are from the energy of two particles with each particle on the same side of the eight-shaped single-particle loop shown in Fig.~\ref{fig:freefermiondeformation}. 
The loops not enclosing $C^*$ are the sum of two energies 
from different sides of the eight-shaped single-particle loops. 
For $\kappa = 0.6$ [see Fig.~\ref{fig:twop_freefermion}\,(bottom)], the two straight lines crossing the origin are from two particles on the same branch of the cross-shaped single-particle spectrum.
On the other hand, the vertical short lines are from two particles on different branches of the cross-shaped single-particle spectrum in Fig.~\ref{fig:freefermiondeformation}.

We also calculate the winding number from the two-particle spectrum, as shown in the right panels of Fig.~\ref{fig:twop_freefermion}. 
For $\kappa = 0.1$ and $\text{Re}\,E > 0$, 
the loop 
that encloses the red reference point in Fig.~\ref{fig:twop_freefermion}\,(top) has the winding number $W = -1$, and the loops 
that enclose the blue and orange reference points 
have the winding number $W = +1$. 
In addition, the purple reference point is enclosed by three outer loops, and therefore the winding number is $W = -3$;
the pink reference point is enclosed by four outer loops and two inner loops and therefore the winding number is 
$W = -2$.
The loops for $\text{Re}\,E < 0$ are characterized by the opposite-sign winding numbers.
Consistently,
the two-particle loops enclosing $C^*$ are from the single-particle spectral loop with the winding number $W = -1$. 
We notice that the presence of the point gap for the two-particle spectrum is the finite-size effect; 
in the infinite-size limit $L \rightarrow \infty$, an infinite number of loops appear and the complex-spectral winding number is no longer well-defined. 
For $\kappa = 0.6$, the loop structure breaks up, and the winding numbers are more complicated.
When the spectral loops break at the transition point $\kappa_c$, the winding number consists of the contributions from all the loops that enclose the reference energy point. 
Therefore, for reference points on the right side of the cross-shaped lines, we have $W<0$; 
for reference points on the upper part of the cross-shaped lines, the winding number can be positive or negative depending on the choice of the reference point.

\section{
Calogero-Sutherland model}\label{sec:CS}

As a prototypical example of interacting integrable models, we consider the Calogero-Sutherland (CS) model~\cite{Calogero_1969, Calogero_1971, Pasquier_1994} for $M$ non-relativistic particles on a circle of length $L$,
\begin{equation}
    H = - 
    \sum_{i=1}^{M}\partial_{x_i}^2 + \left( \frac{\pi}{L} \right)^2\sum_{i<j}^{M}\frac{\beta(\beta -1)}{\sin^2\frac{\pi}{L}(x_i - x_j)}, 
\end{equation}
where $\beta$ describes the strength of the interactions.
The CS model is integrable and solvable using the Bethe ansatz~\cite{Sutherland_1971_JMP1, Sutherland_1971_JMP2, Sutherland_1971_PRA, Sutherland_1972_PRA}. 
The Bethe ansatz equation reads
\begin{equation}
    k_j L 
    + \sum_{l} \Theta(k_j,k_l) = 2\pi I_j,
            \label{eq: CS-BAE}
\end{equation}
with $\Theta(k_j,k_l) \coloneqq \pi(\beta -1) \text{sign}(k_j - k_l)$ and, for the ground state, $I_j \coloneqq j - \frac{M+1}{2}$. 
If we assume 
$k_1 < \cdots < k_M$, we get 
\begin{align}
    \sum_{l=1}^{M}\Theta(k_j,k_l) 
    &= \sum_{l=1}^{M} \pi(\beta -1) \text{sign}(k_j - k_l) \nonumber \\
    &=  -\pi(\beta -1) (2j -M-1).
\end{align}
From Refs.~\cite{Awata_1995,Lapointe_1996,Katsura_2007}, the momentum of excited states of the CS model is given by
\begin{equation}
    k_j = \frac{2\pi}{L}\left(n_j+\beta\Big(j-\frac{M+1}{2}\Big)\right),
\end{equation}
where $n_j$'s are non-negative integers with 
$n_j \leq n_{j+1}$.
We consider the Bethe ansatz equation~\cite{Pavshinkin_2021} with the boost deformation.
The energy of the $j$\,th particle satisfies
\begin{align}
        \varepsilon_j^{\kappa} 
        =\left(k_j+ i\kappa \varepsilon_j^{\kappa}\right)^2
            \label{eq: CS boost equation}
\end{align}

For 
$\kappa = 0$, we recover the nondeformed energy spectrum. 
For $\beta = 0$, the system is bosonic and we recover the same spectrum as the noninteracting system in Sec.~\ref{subsec: free fermion in free space}.
For $\beta = 1$, the system is fermionic and we get the different energies for two particles for the 
$n_j=0$ sector. 
We 
see that for 
$\kappa = 0$,
\begin{align}          \varepsilon_j^{\kappa=0}=
    k_j^2 
    &= \left[\frac{2\pi}{L}\left( 
    -n_j+ \beta \left(\frac{M+1}{2} - j \right) \right)\right]^2,
\end{align}
which is the particle of bare momentum $\beta \left(\frac{M+1}{2} - j \right)$ with 
$-n_j$ partition. 

\begin{figure}[t]
 \centering
  \includegraphics[width=0.49\linewidth]{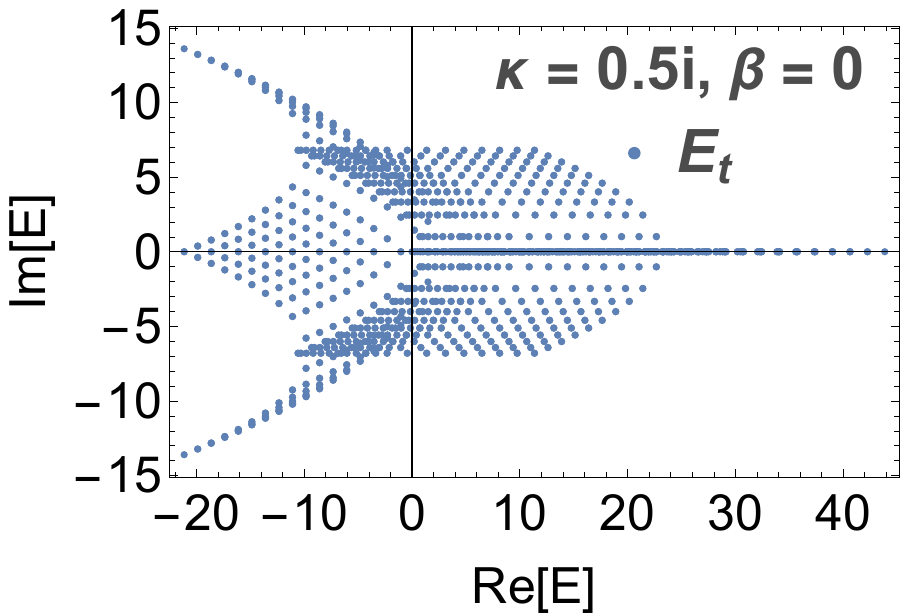}
 \includegraphics[width=0.49\linewidth]{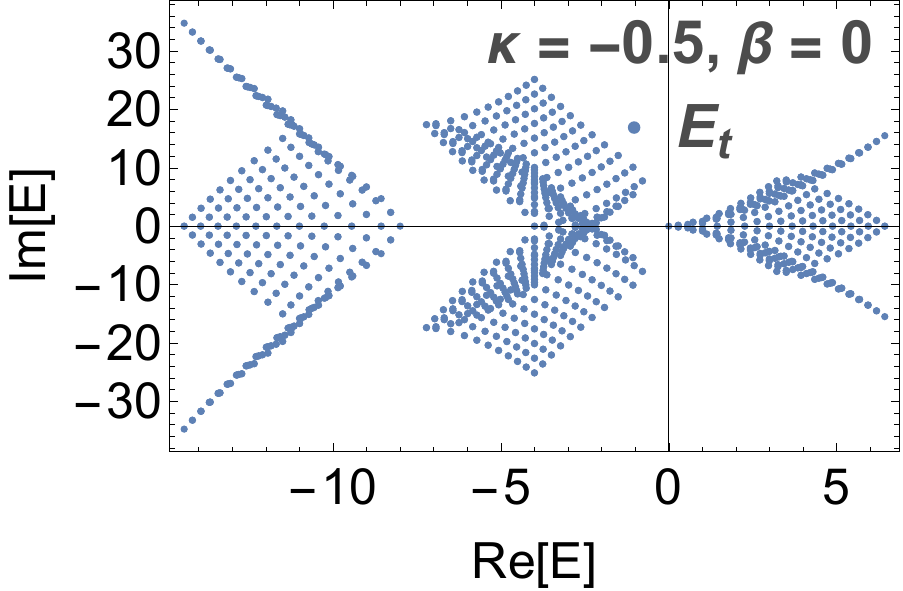}
 \includegraphics[width=0.49\linewidth]{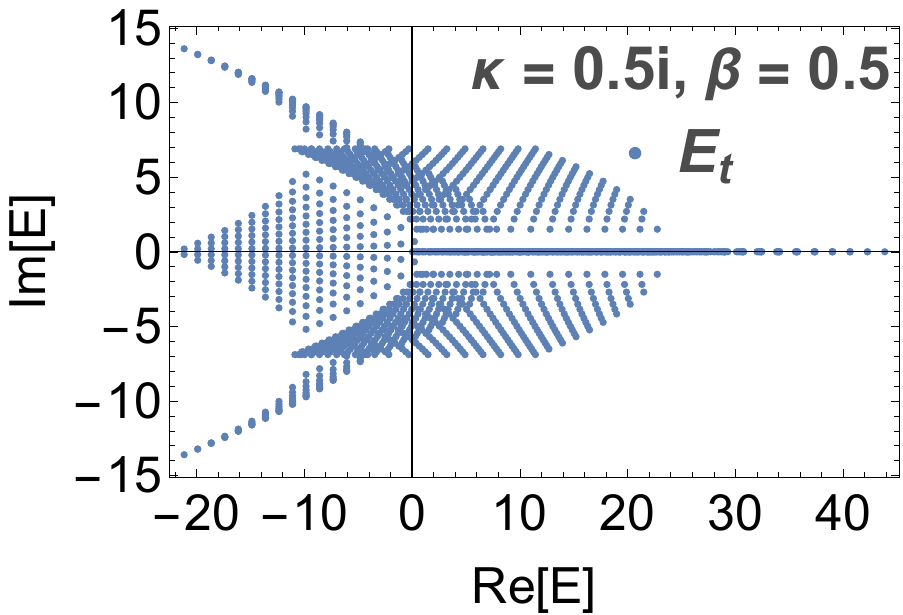}
 \includegraphics[width=0.49\linewidth]{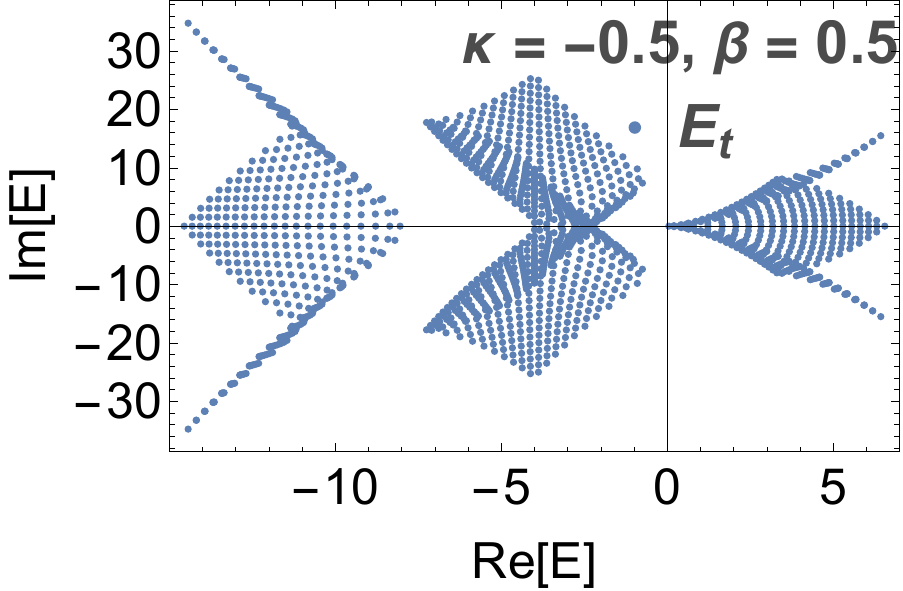}
 \caption{Complex two-particle energy spectrum of the Calogero-Sutherland (CS) model with $\beta = 0$ (top) and $\beta = 0.5$ (bottom) 
 for the (left)~real boost $\kappa = 0.5i$ and  (right)~imaginary boost $\kappa = - 0.5$.
 }
    \label{fig:CS two particle}
\end{figure}

With the boost deformation, the situation is different.
Solving Eq.~(\ref{eq: CS boost equation}),
we calculate the single-particle energy of $n_j \in \mathbb{Z}$ as
\begin{equation}
    \varepsilon_j^{\kappa \pm}(k_j) = - \frac{1}{2\kappa^2}  \Big[1 - 2i\kappa k_j \pm \sqrt{1 - 4i\kappa k_j}\Big],
\end{equation}
and the two-particle energy as $E_t(k_1,k_2) = \varepsilon_1(k_1) + \varepsilon_2(k_2)$ with $\varepsilon_j (k_j) \in {\varepsilon_j^+(k_j)} \cup  {\varepsilon_j^-(k_j)}$.
Figure~\ref{fig:CS two particle} shows the two-particle energy spectrum for the real and imaginary boosts. 
The real-boosted two-particle spectrum with $\kappa = 0.5i$ (see the left plots of Fig.~\ref{fig:CS two particle}) forms a v-shaped spectrum pointing towards the positive $x$ direction. 
On the basis of the single-particle spectrum for the noninteracting case, Fig.~\ref{fig:CS single particle} for $\beta = 0$,
the wings of the v-shape spectrum are from the summation of two particles on each single-particle branch. 
The states clustered between the wings are from two particles on different single-particle branches. 
More states show up in the presence of the interaction (see the left bottom panel of Fig.~\ref{fig:CS two particle} for $\beta = 0.5$). 

In comparison with the two-particle spectrum for the noninteracting case in Fig.~\ref{fig:twop_freefermion}, 
the two-particle spectrum of the CS model with the imaginary boost $\kappa = -0.5$ does not show loop structures but rather forms three clusters of states. 
On the basis of the single-particle spectrum in Fig.~\ref{fig:CS single particle},
the three clusters from the left to the right are from the summation of single-particle energies of the left-left, left-right, and right-right branches. 
In the presence of the interaction, more states show up  and the spectrum gets denser, but no dramatic change is observed. 

\section{
XXZ model}\label{sec:XXZ}

We consider the non-Hermitian boost deformation for the XXZ model 
\begin{equation}
    H = \sum_{i=1}^{L} \Big[ \frac{\cosh{\gamma}}{2} (1+ \sigma_i^z\sigma_{i+1}^z) - \sigma_i^{+}\sigma_{i+1}^{-} - \sigma_i^{-}\sigma_{i+1}^{+} \Big],
\end{equation}
where $\sigma_i^{z}$'s and $\sigma_{i}^{\pm}$'s are Pauli operators, and 
$\gamma$
is the parameter that controls the many-body interactions.
The Bethe ansatz equation is given by~\cite{takahashi_1999, Albertini_1997}
\begin{equation}
    \Bigg[\frac{\sinh(\frac{\gamma}{2}+\frac{i\alpha_k}{2})}{\sinh(\frac{\gamma}{2}-\frac{i\alpha_k}{2})}\Bigg]^L = (-1)^{N+1} \prod_{l=1}^N \frac{\sinh(\gamma + \frac{i}{2}(\alpha_k - \alpha_l))}{\sinh(\gamma - \frac{i}{2}(\alpha_k - \alpha_l))},
\end{equation}
where $\alpha_k$ labels the rapidity of particle $k$, 
and the ground state energy is given by
\begin{equation}
    E = L\cosh\gamma - \sum_{k=1}^N \frac{2\sinh^2 \gamma}{\cosh\gamma - \cos\alpha_k}.
\end{equation}
To simplify calculations, we define a new variable 
\begin{equation}
    z_k \coloneqq \frac{\sinh(\frac{\gamma}{2}+\frac{i\alpha_k}{2})}{\sinh(\frac{\gamma}{2}-\frac{i\alpha_k}{2})},
\end{equation}
and then the Bethe ansatz equation reduces to 
\begin{equation}
    z_k^L = (-1)^N\prod_{l=1}^N \frac{1+z_k z_l + 2\cosh\gamma z_k}{1+z_k z_l + 2\cosh\gamma z_l}.
\end{equation}
In particular, the energy of each particle is 
\begin{equation}
\varepsilon_k = -  \frac{2\sinh^2 \gamma}{\cosh\gamma - \cos\alpha_k} = -( z_k + z_k^{-1} +2\cosh\gamma ) .
\end{equation}

\begin{figure}[t]
 \centering
  \includegraphics[width=0.48\linewidth]{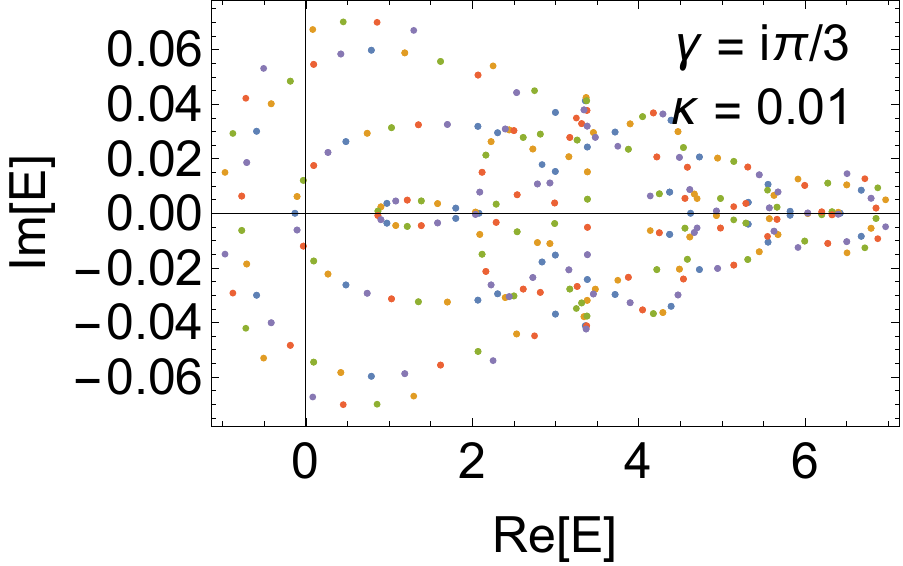}
 \includegraphics[width=0.48\linewidth]{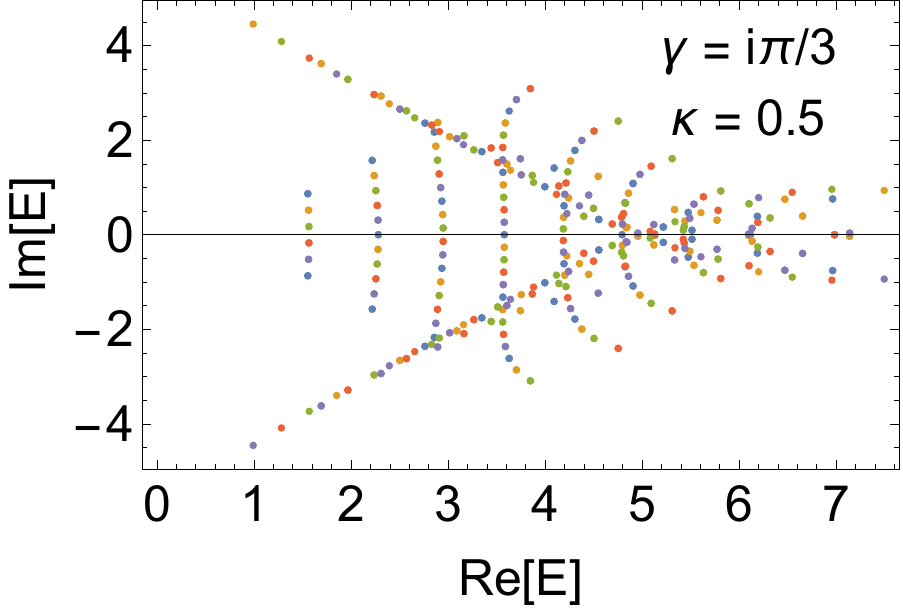}
 \includegraphics[width=0.48\linewidth]{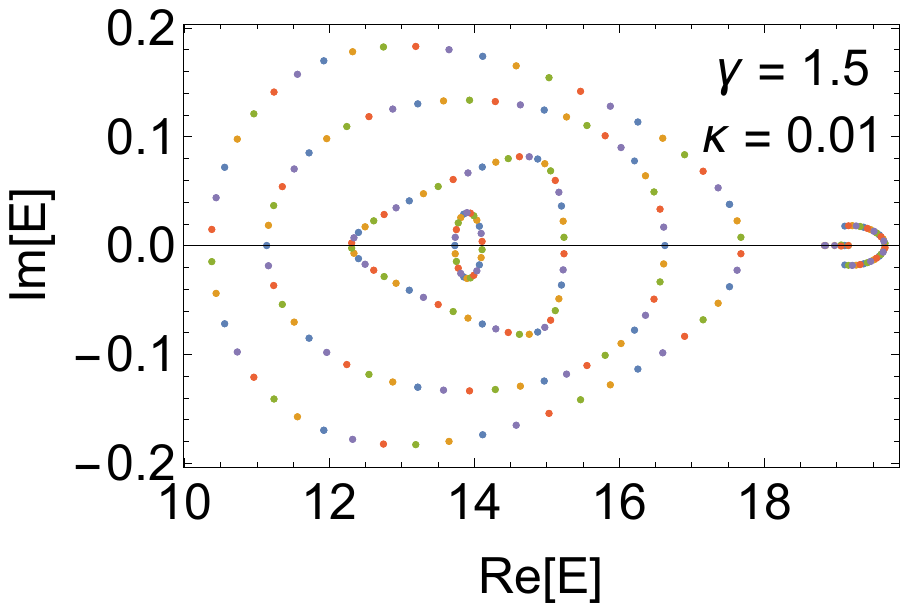}
 \includegraphics[width=0.48\linewidth]{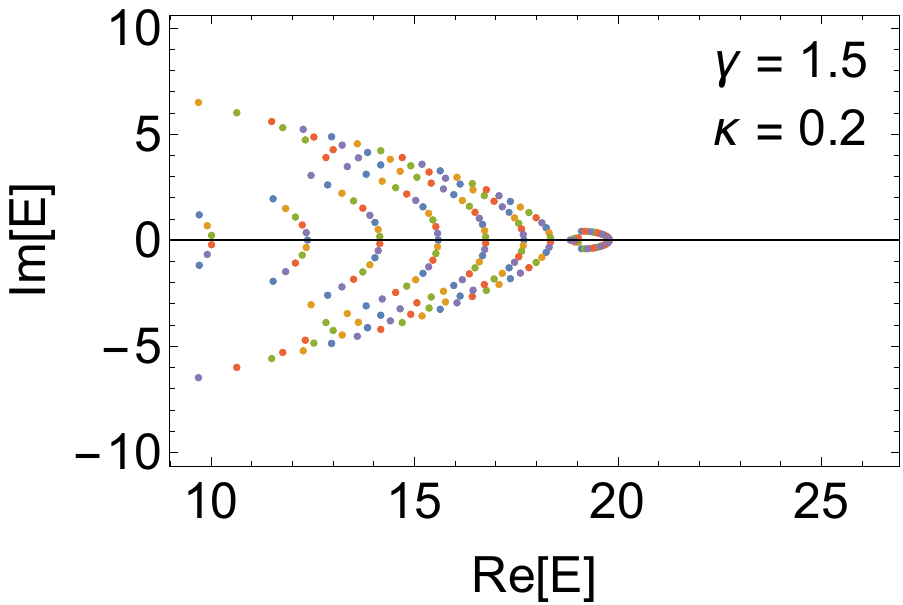}
 \caption{Complex two-particle energy spectrum of the XXZ model with $L=10$ and (top)~$\gamma  =i\pi/3$ and (bottom)~$\gamma = 1.5$ under the imaginary boost deformation of $\kappa = 0.01, 0.2, 0.5$. The phase twist is chosen to be $\phi = \{0, \frac{2\pi}{5},\frac{4\pi}{5},\frac{6\pi}{5},\frac{8\pi}{5}\}$.
 }
    \label{fig:xxz_twoparticle}
\end{figure}

For the imaginary boost deformation, following Eq.~(\ref{eq:momentum boost}), we introduce 
\begin{equation}
p_k \rightarrow p_k - i\kappa \varepsilon_k 
\end{equation}
with the identification 
$z_k = e^{i p_k}$ from Ref.~\cite{takahashi_1999}.
The deformed Bethe ansatz equation reads
\begin{align}
        &z_k^L e^{-   L\kappa(z_k + z_k^{-1} + 2\cosh\gamma)} \nonumber \\
        &\qquad\qquad= (-1)^N
        \prod_{l=1}^N \frac{1+z_kz_l + 2\cosh\gamma z_k}{1+z_kz_l + 2\cosh\gamma z_l}.
\end{align}
To emulate a large system size, we also consider the model with a U(1) phase twist with twist angle $\phi$, resulting in the Bethe ansatz equation 
\begin{align}
        &z_k^L e^{-   L\kappa (z_k + z_k^{-1} + 2\cosh\gamma)} \nonumber \\
        &\qquad\quad= e^{-i\phi}(-1)^N
        \prod_{l=1}^N \frac{1+z_kz_l + 2\cosh\gamma z_k}{1+z_kz_l + 2\cosh\gamma z_l}.
\label{eq:bae boost}
\end{align}
We numerically solve the Bethe ansatz equation for $N=2$ and obtain the two-particle complex spectrum for the weak and strong interacting cases of $\gamma = i\pi/3$ and $\gamma = 1.5$, as shown in Fig.~\ref{fig:xxz_twoparticle}. 
Similarly to the free-fermion case in Sec.~\ref{subsec: free fermion - two particle}, we focus only on some regions of the complex spectrum where the Bethe ansatz equation is solved consistently.

For the weakly interacting cases, the spectrum is similar to the free fermion case in Fig.~\ref{fig:twop_freefermion}.
In fact, there are multiple loops, and some of them cross themselves for small $\kappa$;
the loops break up into some pieces of arcs as $\kappa$ increases. 
Short vertical curves are also observed as in the case of free fermion. 
The weak interaction results in only half of the cross-shaped structure appearing in the free fermion case due to interaction. 
The v-shaped curves are formed by quasiparticles on the same branch of single-particle energy, as is shown in Fig.~\ref{fig:freefermiondeformation}. 
The vertical lines are formed by two quasiparticles on different branches. 

For strongly interacting cases, the complex spectrum forms multiple loops and there is a small cluster of states away from the main loops for small $\kappa < 0.1$, and the loops are pushed to the $\operatorname{Re}\,E$ direction and break up into arcs as $\kappa$ increases (see the bottom panels of Fig.~\ref{fig:xxz_twoparticle}).
This behavior is similar to the interacting Hatano-Nelson model~\cite{Zhang_Neupert_2022, Kawabata_Shiozaki_Ryu_2022}. 
The small cluster of states that is not observed for the weakly interacting cases should be due to the strong interaction effect. 
In comparison with the two-particle spectrum of non-interacting fermions in Fig.~\ref{fig:twop_freefermion},
the transition from the closed loops to the open curves happens earlier for the interacting XXZ model and the transition is earlier as the interaction is stronger. 
After the transition, the small cluster of states gets closer to the main cluster of states. 

\section{Conclusion}
    \label{sec:conclusion}

In this work, we extended the original idea of Hatano and Nelson~\cite{Hatano_Nelson_1996, Hatano_Nelson_1997} to a new class of non-Hermitian Hamiltonians using the imaginary boost deformation for integrable systems.
The imaginary boost deformation can be viewed as a scale transformation that generates non-Hermitian Hamiltonians as an analog of the Galilean transformation used in the Hatano-Nelson model. 
For integrable systems, the imaginary boost deformation complexifies the spectral parameter. 
We identified that such an imaginary boost deformation can be formulated as an energy-twisted boundary condition using a chiral Dirac fermion. 

We studied our imaginary boost deformation for several integrable models with and without many-body interactions.
We implemented the imaginary boost deformation in free fermion systems 
in the continuum 
and on a lattice.
We observed unique complex spectra with non-trivial winding numbers. 
The two branches of complex spectra emerge from infinity as soon as the deformation is turned on. 
The similarity transformation was performed in the open boundary conditions to reveal the non-Hermitian skin effect.
We showed the two-particle spectrum of the Calogero-Sutherland model and the XXZ model and observed the non-Hermitian many-body interaction effect even for a small boost parameter. 
The spectrum breaks into arcs at large boost parameters similar to the non-interacting models.
Our results mathematically provide a nontrivial manner to generate non-Hermitian integrable systems and physically provide a new perspective on spacetime transformations.
Additionally, our approach should be relevant to a large class of integrable open quantum systems.

The origin of the energy emergent from infinity needs further study and characterization. 
Extending the imaginary boost deformation to quantum many-body systems in higher dimensions would be interesting to study.
It is also worthwhile to study transport properties due to the imaginary boost deformation in a similar manner to the thermal Drude weight for the real boost deformation~\cite{Nakai_Guo_Ryu_2022}.
In fact, the imaginary gauge field in the Hatano-Nelson model is closely related to the current and the delocalization of wave functions~\cite{Hatano_Nelson_1996, Hatano_Nelson_1997}.
Another direction to pursue is to generalize the imaginary boost parameter to other sorts of deformations such as bilinear deformations~\cite{Bargheer_2009}. 
The dynamical properties including the spectral form factor of non-Hermitian deformations are also investigated in the recent work~\cite{Matsoukas-Roubeas_2022}. 
Similarly, the dynamical properties of the imaginary boosted systems are also worth studying in future works.

\section*{Acknowledgements}
We thank Hosho Katsura and Ken Shiozaki for helpful discussions.
K.K. is supported by the Japan Society for the Promotion of Science (JSPS) through the Overseas Research Fellowship.
R.N. is supported by JSPS KAKENHI Grant No. JP17K17604 and JST CREST Grant No. JPMJCR18T2.
S.R. is supported by the National Science Foundation under award number DMR-2001181, and by a Simons Investigator Grant from the Simons Foundation (Award Number: 566116).
This work is supported by the Gordon and Betty Moore Foundation through Grant GBMF8685 toward the Princeton theory program.

\appendix

\section{Boost deformation and similarity transformation}
    \label{appendix: boost}

In this section, we show how the boost deformation in Eq.~(\ref{eq:boostdeformation}) is related to the similarity transformation in Eq.~(\ref{eq:similaritytransformation_continuousfreefermion}) via Eq.~(\ref{eq:similaritytransformation_integrable}).
We also discuss the similarity transformation for a lattice free fermion.

\subsection{Continuum free fermion}
\label{sec:similaritytransformation_continuous}

To derive the similarity transformation from Eq.~(\ref{eq:similaritytransformation_integrable}), we need to know the boost operator and hence the deformed Hamiltonian.
Let us derive the boost deformation of the continuum free fermion 
in Sec.~\ref{subsec: free fermion in free space}.
As we see from the definition in Eq.~(\ref{eq:boostdeformation}), the boost deformation generates higher derivatives.
Hence, we use an ansatz 
\begin{align}
    \mathcal{H}(\kappa)=\sum_{n=0}A_n(\kappa)\partial_x^n 
    \label{eq:ansatzhamiltonian_continuousfreefermion}
\end{align}
and solve equations order by order in $\kappa$ with the initial condition $\mathcal{H}(\kappa=0)=-\partial_x^2$.
Substituting Eq.~(\ref{eq:ansatzhamiltonian_continuousfreefermion}) into Eq.~(\ref{eq:boostdeformation}), we obtain a series of equations as 
\begin{align}
    &\frac{dA_0}{d\kappa}=-A_0A_1,\\
    &\frac{dA_1}{d\kappa}=-A_1A_1-2A_0A_2,\\
    &\frac{dA_2}{d\kappa}=-A_1A_2-2A_1A_2-3A_3A_0,
\end{align}
and so on.
As the boost deformation generates higher derivatives than quadratic, we assume $A_0(\kappa)=A_1(\kappa)=0$.
With the initial condition $A_n(\kappa=0)=-\delta_{n,2}$, the solution is
\begin{align}
    H(\kappa)=(-i\partial_x)^2+2i\kappa(-i\partial_x)^3-5\kappa^2(-i\partial_x)^4+\cdots,
\end{align}
which is equivalent to the series expansion of $\varepsilon_k^{-}$ in Eq.~(\ref{eq:deformedspectrum_continuousfreefermion}) by replacing $k_0$ by $-i\partial_x$.

Then, we solve Eq.~(\ref{eq:similaritytransformation_integrable}), which reads
\begin{align}
    \frac{d\psi(x;\kappa)}{d\kappa}=-\frac{1 - 2i \kappa(-i\partial_x) - \sqrt{1 -  4i\kappa(-i\partial_x)}}{2\kappa^2}\psi(x;\kappa).
\end{align}
We use an ansatz $\psi(\kappa)=e^{if(\kappa)x}$.
The equation for $f(\kappa)$ reduces to 
\begin{align}
    \frac{df}{d\kappa}=\frac{i(1 - 2i \kappa f - \sqrt{1 -  4i\kappa f})}{2\kappa^2}.
\end{align}
By introducing $F \coloneqq (1-4i\kappa f)^{1/2}$, the equation becomes $dF/d\kappa=(F-1)/\kappa$ and the solution is $F=a\kappa+1\,(a\in\mathbb{C})$.
When the initial state is $e^{if(\kappa=0)x}=e^{ik_0x}$, the deformed state is
\begin{align}
    \psi(x;\kappa)=\exp\left(ik_0x+\kappa k_0^2 x\right),
    \label{eq:deformedwavefunction_continuousfreefermion}
\end{align}
which is consistent with the similarity transformation in Eq.~(\ref{eq:similaritytransformation_continuousfreefermion}).

Notice that by leaving the wave function $e^{ik_0x}$ unchanged, the eigenenergy is deformed as $\varepsilon_k^{-}$ in Eq.~(\ref{eq:deformedspectrum_continuousfreefermion}),
while by deforming the wave function as Eq.~(\ref{eq:deformedwavefunction_continuousfreefermion}), the eigenenergy $\varepsilon=k_0^2$ is unchanged as was also shown in Sec.~\ref{sec:ABA}.
The fact that the initial wave function is still an eigenfunction of the deformed Hamiltonian 
seems to be a coincidence unique to the continuum free fermion model.

\subsection{Lattice free fermion}
\label{sec:similaritytransformation_lattice}

Let us consider the similarity transformation for the lattice free fermion under the open boundary conditions studied in Sec.~\ref{sec: free fermion on lattice OBC}. 
The imaginary boost deformation in Eq.~(\ref{eq:boostdeformation_latticefermion}) is written in the matrix form as
\begin{align}
    \frac{dT}{d\kappa}=\frac{1}{2}[X,T^2],
\end{align}
where $[T(\kappa)]_{xy}=t_{xy}(\kappa)$ is the Hamiltonian matrix and $(X)_{xy}=x\delta_{xy}$.
We introduce the similarity transformation $T(\kappa)=U(\kappa)T_0U^{-1}(\kappa)$ to make a connection between the deformed and original Hamiltonian matrices, with $T_0=T(\kappa=0)$.
The imaginary boost deformation becomes
\begin{align}
    \left[U^{-1}\frac{dU}{d\kappa},T_0\right]=\frac{1}{2}\left[\left\{U^{-1}XU,T_0\right\},T_0\right],
\end{align}
which is satisfied for
\begin{align}
    U^{-1}\frac{dU}{d\kappa}=\frac{1}{2}\left\{U^{-1}XU,T_0\right\}.
    \label{eq:equationofu_latticefermion}
\end{align}
We solve this equation order by order in $\kappa$ by using an expansion $U(\kappa)=\sum_{n=0}\kappa^nU_n/n!$.
The solution is 
\begin{align}
    U(\kappa)=I+\frac{\kappa}{2}\{X,T_0\}+\frac{\kappa^2}{8}\left(\{X,T_0\}^2+[X^2,T_0^2]\right)+O(\kappa^3).
\end{align}
For the initial condition $(T_0)_{xy}=\delta_{x,y+1}+\delta_{x,y-1}$,
the eigenfunction without the boost deformation is $\psi_k(x)\propto\sin kx\,(x=\pi n/(L+1),\,n\in[1,L])$, which satisfies the boundary conditions $\psi_n(0)=\psi_n(L+1)=0$,
and the corresponding eigenenergy is $\varepsilon_k=2\cos k$.

\begin{figure}[t]
    \centering
    \includegraphics[width=\columnwidth]{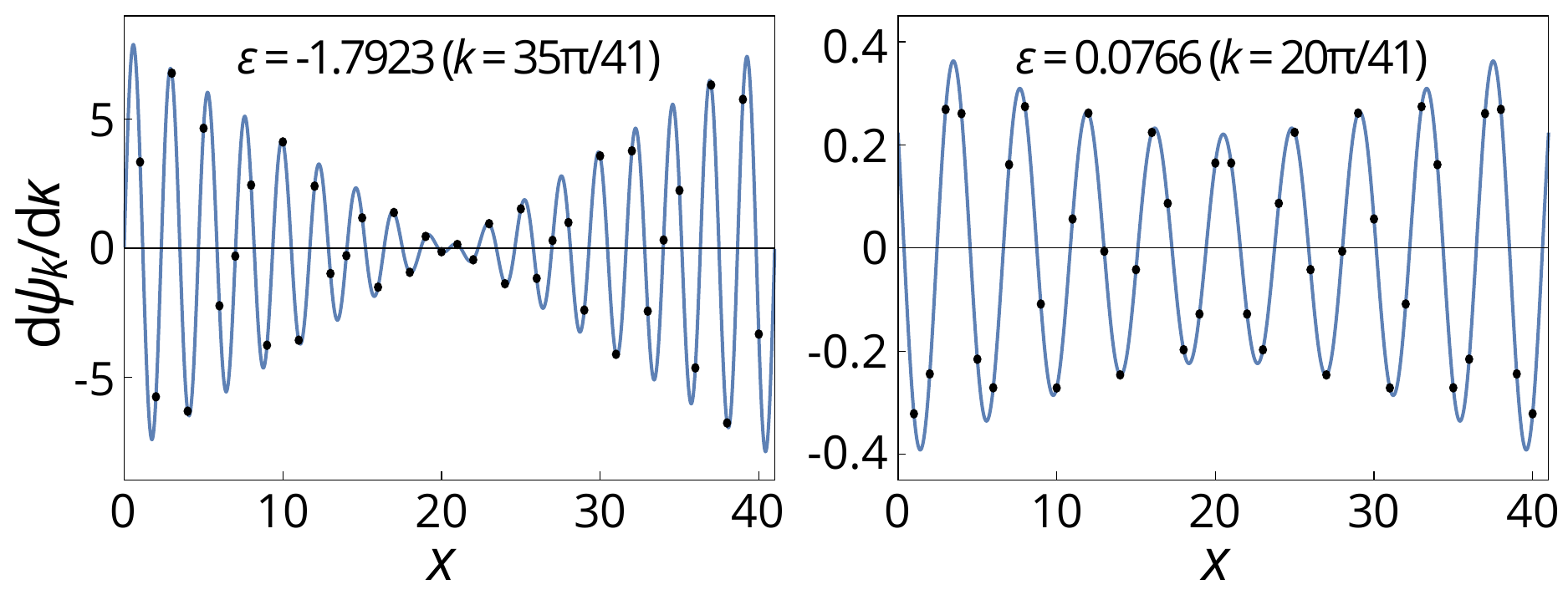}
    \caption{Derivative of eigenfunctions of the lattice free fermion of length $L=40$ with respect to the boost parameter $\kappa$ evaluated by numerical (dots) and analytical (solid lines) methods for
    $k=35\pi/41$ (left) and $k=20\pi/41$ (right).
        \label{fig:similarity_lattice}
    }
\end{figure}

Since we have
\begin{align}
    \frac{1}{2}\{X,T_0\}
    =
    \begin{pmatrix}
        0 & 3/2 & 0 & \\
        3/2 & 0 & 5/2 & \\
        0 & 5/2 & 0 & \\
        &&&\ddots
    \end{pmatrix},
    \label{eq:similarityfirstorder_latticefermion}
\end{align}
the deformed eigenfunction up to linear in $\kappa$ is
\begin{align}
    U\psi_k\simeq
    \begin{pmatrix}
        (1+2\kappa\cos k)\sin k +  \kappa\cos k\sin k\\
        (1+4\kappa\cos k)\sin 2k + \kappa\cos 2k\sin k\\
        (1+6\kappa\cos k)\sin 3k + \kappa\cos 3k\sin k\\
        \vdots
    \end{pmatrix}.
    \label{eq:similaritytransformaiton_latticefermion_linear}
\end{align}
For $k\simeq 0$ or $\pi$, the first term in each row is dominant and the similarity transformation is consistent with $\psi_k(x)\simeq e^{\kappa \varepsilon_k x}\sin kx$ 
for the continuum free fermion in Eq.~(\ref{eq:similaritytransformation_continuousfreefermion}).
On the other hand, for $k\simeq\pi/2$, Eq.~(\ref{eq:similaritytransformaiton_latticefermion_linear}) implies that the similarity transformation cannot be written as Eq.~(\ref{eq:similaritytransformation_continuousfreefermion}).

In Fig.~\ref{fig:similarity_lattice}, we compare the derivative of the eigenfunction evaluated numerically for the deformed Hamiltonian in Eq.~(\ref{eq:deformedhamiltonian_latticefreefermion}) with that obtained from the analytical expression in Eq.~(\ref{eq:similaritytransformaiton_latticefermion_linear}). 
We see that these two agree quite well.
Notice that, in the analytical expression, the first order term in Eq.~(\ref{eq:similarityfirstorder_latticefermion}) is shifted by $-(L+1)T_0/2$ 
since this is an ambiguity in deriving Eq.~(\ref{eq:equationofu_latticefermion}).
For $k\simeq 0$ or $\pi$, the first term of Eq.~(\ref{eq:similaritytransformaiton_latticefermion_linear}) dominates and the derivative is roughly proportional to $-(L+1)/2+x$ [Fig.~\ref{fig:similarity_lattice} (left)],
while such a feature cannot be seen for $k\simeq \pi/2$ [Fig.~\ref{fig:similarity_lattice} (right)].

\subsection{Numerical instability}

\begin{figure}[t]
  \centering
  \includegraphics[width=70mm]{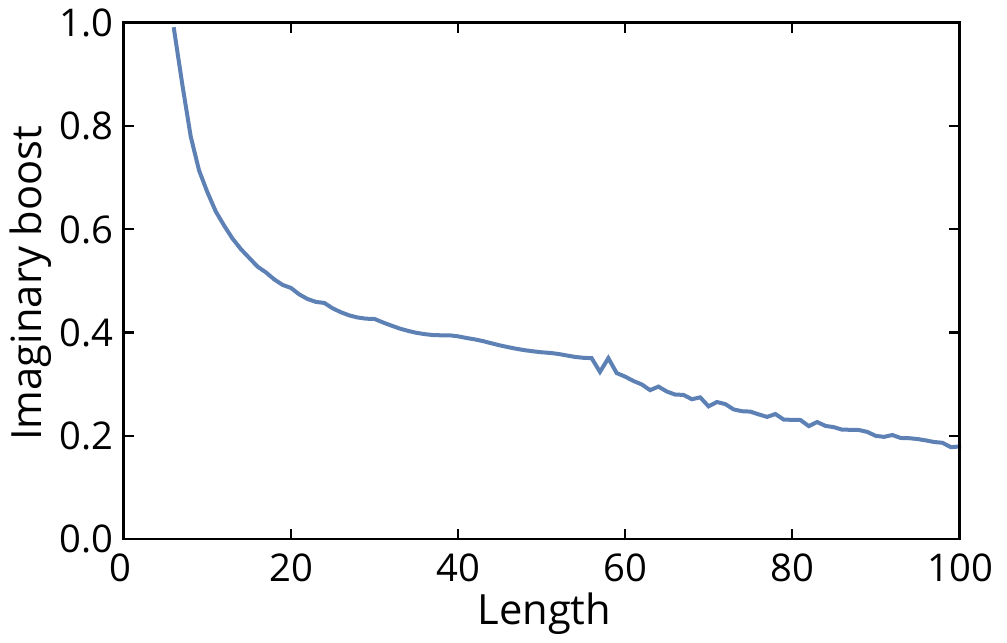}
  \caption{Imaginary boost parameter at which the spectrum of the lattice free fermion deviates from the real axis.}
    \label{fig:numericalinstability}
\end{figure}

According to the similarity transformation, the energy spectrum of the lattice free fermion in Eq.~(\ref{eq:boostdeformation_latticefermion}) under the open boundary conditions is unchanged and remains real-valued by the imaginary boost deformation.
However, the numerically evaluated Hamiltonian matrix can have a complex spectrum when the imaginary boost parameter becomes larger because of the instability of the numerical calculations.
The critical parameter from which the spectrum becomes complex is numerically evaluated in Fig.~\ref{fig:numericalinstability}.
We see that the numerical instability 
becomes more serious as the system size is larger.
Notice that the numerical instability is not relevant to the spectral transition point in Eq.~(\ref{eq: kappa critical}) under the periodic boundary conditions.

\bibliography{reference}

\end{document}